\documentclass[seceq]{ptptex}
\title{Nonintegrability and fluctuation by symmetry violation\\ in perturbed harmonic oscillator systems}

\author{Shigeyasu \textsc{FUJIWARA}}

\inst{General Arts, Hiroshima National College of Maritime
Technology, 4272-1, Higashino, Oosakikamishima-cho, Toyota-gun, Hiroshima
725-0231, Japan}

\abst{The violation of symmetry between the time series of elongation and contraction rates of phase-space point spacings is studied to examine the chaos in perturbed harmonic oscillator systems.
A transition from integrability to nonintegrability is quantitatively evaluated by introducing a correlation coefficient for the degree of symmetry violation.
A feature of the correlation coefficient is discussed, suggesting a possible advantage over the Lyapunov exponent.
The relationship between a fluctuation property based on symmetry violation and nonintegrability is also discussed.}

\begin{document}

\maketitle
\section{Introduction}

We demonstrate the characterization of chaos by exploiting a correlation coefficient for a rate of elongations and contractions of consecutive phase-space point spacings.
In the case of an integrable orbit on a Poincar$\acute{\textrm e}$ section plane\cite{rf:1,rf:2}, it is found that a degree of symmetry occurs between the elongation and contraction rates of the consecutive phase-space point spacings.
When this symmetry occurs, the absolute value of the correlation coefficient becomes $1.0$.
On the other hand, the symmetry of the rates of elongation and contraction is violated when integration becomes impossible, and the absolute values of correlation coefficients are close to zero.
This analysis provides us with a new method of considering how the nonintegrabilities depend on the degree of symmetry violation between the elongations and contractions of consecutive phase-space
point spacings in perturbed harmonic oscillator systems.
We compare the information obtained from the absolute value of the correlation coefficient with that obtained from the Lyapunov exponent\cite{rf:1,rf:2} and show the advantage of using the correlation coefficient.
Moreover, we show that a fluctuation property derived from the consecutive phase-space point spacings follows the shape of a non-Gaussian distribution function as the degree of symmetry violation increases.

\

\section{Consecutive phase-space point spacings}

To begin with, let us define $l_{n}$. 
The spacing of consecutive phase-space points on the Poincar$\acute{\textrm e}$ section plane is defined as $l_{n}\equiv\sqrt{(q_{n+1}-q_{n})^{2}+(p_{n+1}-p_{n})^{2}}$. 
Here, $n$ enumerates a set of points on the Poincar$\acute{\textrm {e}}$ section plane. 
The scaled spacings of $l_{n}$ is defined as $L_{n}\equiv {l_{n}}\cdot({\bar{l}_{n}})^{-1}$, 
where the bar denotes a "spacing average'' over all spacings. 
The spacings were scaled to make $\bar{L}_{n}=1.0$. 
The following modified nondimensional ${\textit {SU}} $(3) Hamiltonian\cite{rf:3,rf:4} was adopted in analyzing $L_{n}$:
\begin{eqnarray}
H(q^{(1)},p^{(1)};q^{(2)},p^{(2)})=
2{\cal E}^{(0)}r^{2}+{\cal E}^{(1)}\frac{({q^{(1)}}^{2}+{p^{(1)}}^{2})}{2}+{\cal E}^{(2)}\frac{({q^{(2)}}^{2}+{p^{(2)}}^{2})}{2}\nonumber\\
+V^{(1)}\frac{2({\cal N}-1)}{{\cal N}}r^{2}({q^{(1)}}^{2}-{p^{(1)}}^{2})
+V^{(2)}\frac{2({\cal N}-1)}{{\cal N}}r^{2}({q^{(2)}}^{2}-{p^{(2)}}^{2})\hspace{1mm},
\end{eqnarray}
where
$r\equiv\frac{1}{2}\sqrt{2{\cal N}-{q^{(1)}}^{2}-{p^{(1)}}^{2}-{q^{(2)}}^{2}-{p^{(2)}}^{2}}$,
and
$V^{(1)}$ and $V^{(2)}$ denote the strength of perturbation interaction in each degree of freedom.
 The system with the Hamiltonian in Eq. (2.1) consists of three single-particle levels with energies ${\cal E}^{(0)}$, ${\cal E}^{(1)}$, and ${\cal E}^{(2)}$, where each level has ${\cal N}$-fold degeneracy.
All subsequent analysis assumes that $V^{(1)}=V^{(2)}=V$.
In our numerical calculation, we used ${\cal N}=30$, and the single-particle energies were ${\cal E}^{(0)}=0$, ${\cal E}^{(1)}=1$, and ${\cal E}^{(2)}=2$.
The position at time $t$ is written as $q^{(1)}_{n}$, and the position at time $t+\epsilon$ is written as $q^{(1)}_{n+1}$, where $\epsilon$ represents the time step of our simulation.

The succession of mappings with the Hamiltonian in Eq. (2.1) is approximated as
\begin{eqnarray}
q^{(1)}_{n+1}&=&q^{(1)}_{n}+\frac{\partial H(q^{(1)}_{n},p^{(1)}_{n};q^{(2)}_{n},p^{(2)}_{n})}{\partial p^{(1)}_{n}}\epsilon\hspace{1mm},\nonumber\\
p^{(1)}_{n+1}&=&p^{(1)}_{n}-\frac{\partial
H(q^{(1)}_{n},p^{(1)}_{n};q^{(2)}_{n},p^{(2)}_{n})}{\partial
q^{(1)}_{n}}\epsilon\hspace{1mm},\nonumber\\
q^{(2)}_{n+1}&=&q^{(2)}_{n}+\frac{\partial H(q^{(1)}_{n},p^{(1)}_{n};q^{(2)}_{n},p^{(2)}_{n})}{\partial p^{(2)}_{n}}\epsilon\hspace{1mm},\nonumber\\
p^{(2)}_{n+1}&=&p^{(2)}_{n}-\frac{\partial
H(q^{(1)}_{n},p^{(1)}_{n};q^{(2)}_{n},p^{(2)}_{n})}{\partial
q^{(2)}_{n}}\epsilon\hspace{1mm}.
\end{eqnarray}
Hereafter the discussion focuses on $q^{(1)}$ and $p^{(1)}$, which are written as $q$ and $p$ for simplicity.

\begin{figure}[htb]
 \begin{center}
  \begin{minipage}{36mm}
   \begin{center}
    \unitlength=2mm
     \special{epsfile=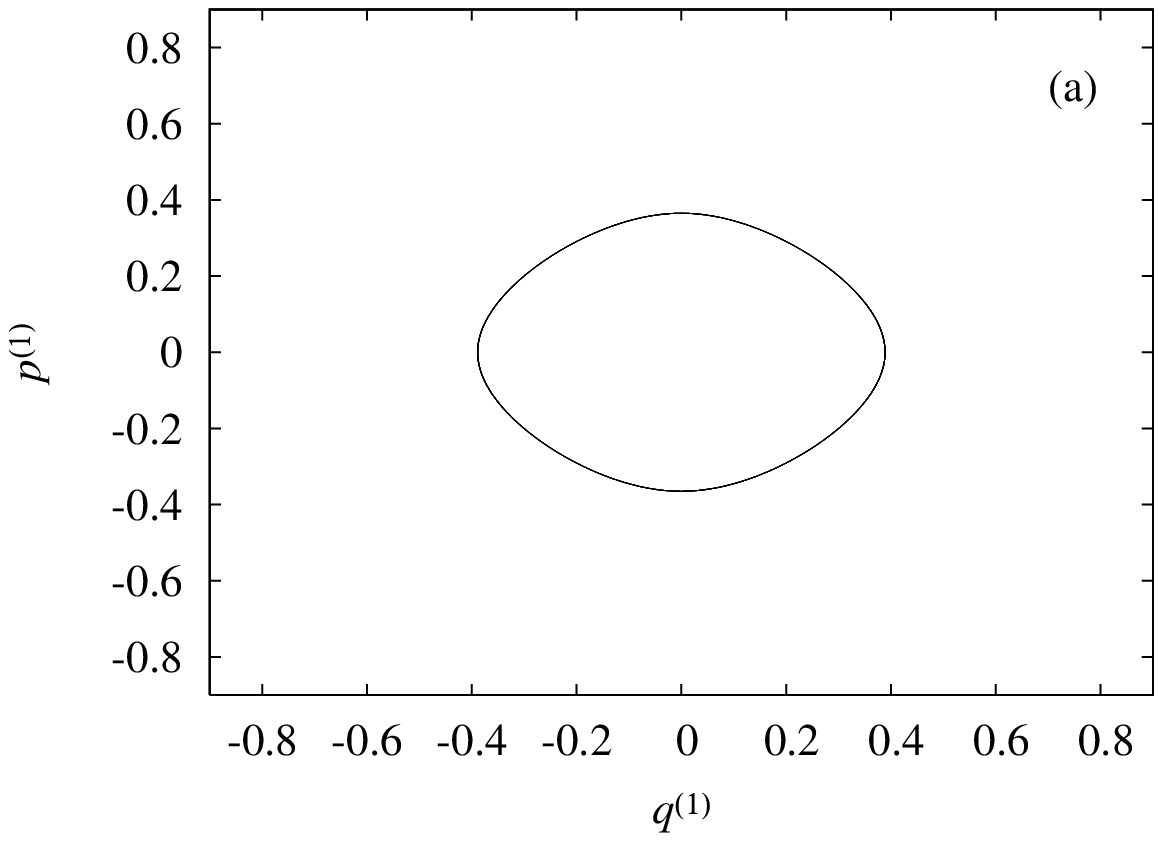 hscale=0.37 vscale=0.53}
      \vspace{51mm}
        \end{center}
         \end{minipage}
          \hspace{8mm}
         \begin{minipage}{36mm}
        \begin{center}
       \unitlength=2mm
      \special{epsfile=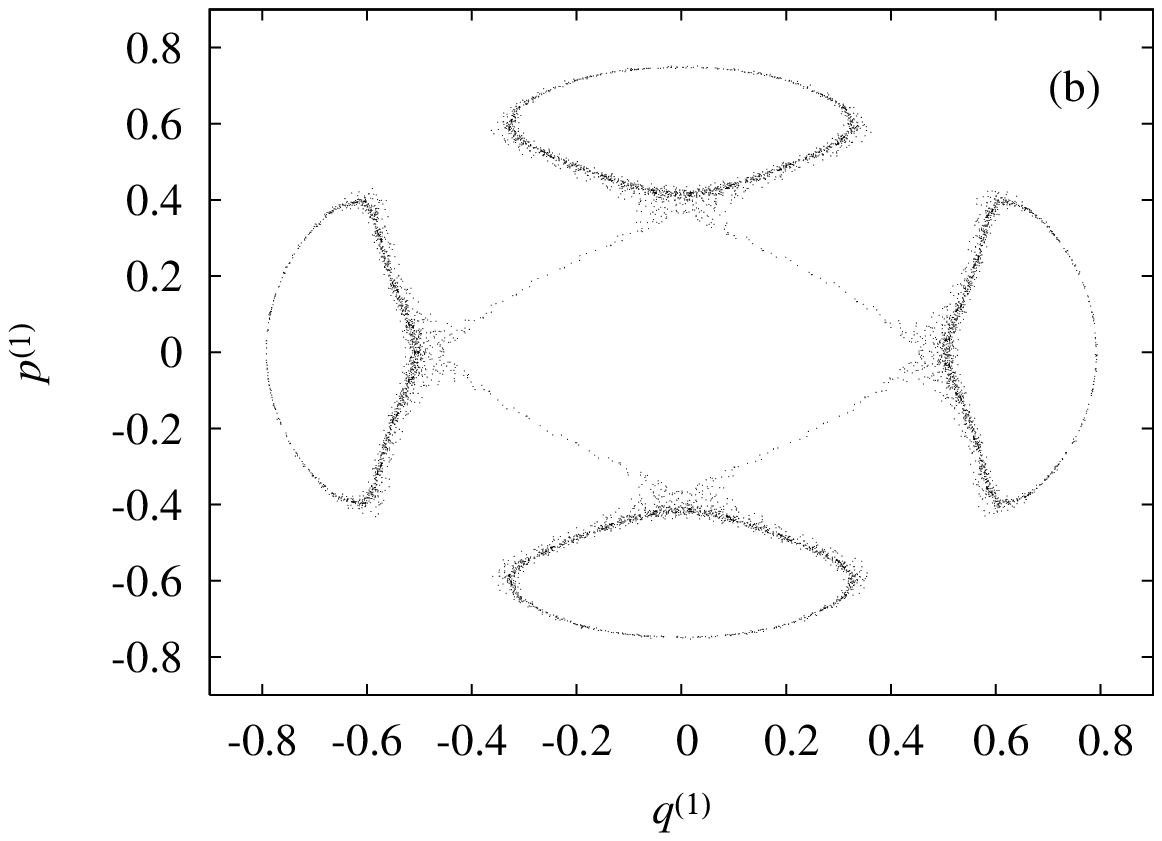 hscale=0.37 vscale=0.53}
     \vspace{51mm}
   \end{center}
  \end{minipage}
          \hspace{8mm}
         \begin{minipage}{36mm}
        \begin{center}
       \unitlength=2mm
      \special{epsfile=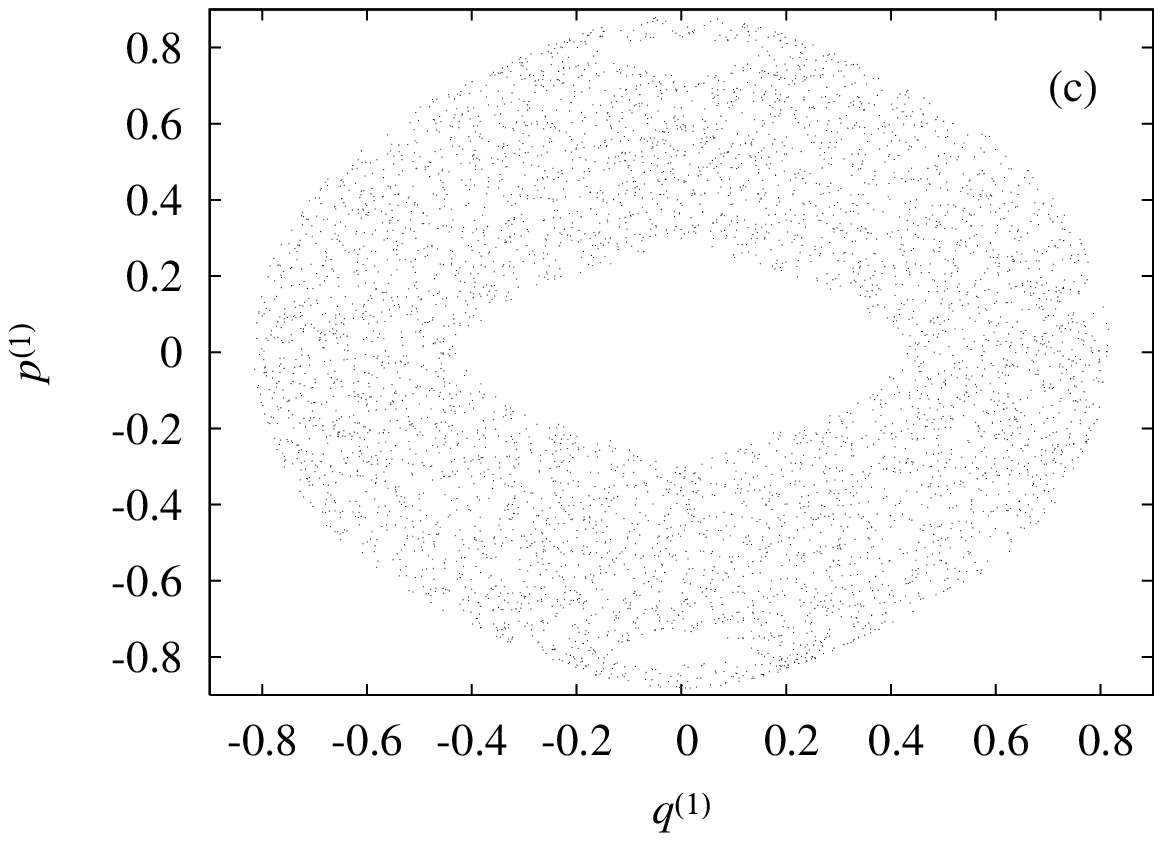 hscale=0.37 vscale=0.53}
     \vspace{51mm}
   \end{center}
  \end{minipage}
 \end{center}
       \caption{Poincar$\acute{\textrm{e}}$ section plane for the ${\textit {SU}}$(3)
Hamiltonian with (a) $V=-0.009$, (b) $V=-0.045$, and (c) $V=-0.110$. }
         \end{figure}

If the interaction $V$ is sufficiently small,
the Poincar$\acute{\textrm{e}}$ section map turns out to be closed curves, as shown in Fig. 1(a). 
Many orbits form invariant curves in this map, like Fig.\hspace{0.8mm}1(a), as guaranteed by the KAM theorem. \cite{rf:1,rf:2} 
In this weak-interaction case, the time dependence of $L_{n}$ turns out to be periodic, as shown in Fig.\hspace{0.8mm}2(a) and Fig.\hspace{0.8mm}3(a). 
The alternating behavior of $L_n$ for this case is shown in more detail in Fig.\hspace{0.8mm}2(d). 
The value of $L_{n}$ is distributed in the range $0.943$ to $1.03$, giving a distribution width of $0.087$.
As is clearly seen from Fig.\hspace{0.8mm}3(a), the Fourier components\cite{rf:5} expressing the
mixing effects from other periodic orbits are small in the case of a small interaction $V=-0.009$.
Comparing Fig.\hspace{0.8mm}3(a) with Fig.\hspace{0.8mm}3(b), one can see that the peak of the Fourier coefficients becomes continuous when the interaction strength increases from  $V=-0.009$ to  $V=-0.045$. 
As the interaction strength increases further, the chaotic region begins to spread out from the hyperbolic fixed points. This is shown in Fig.\hspace{0.8mm}1(c).
In this chaotic case, the value of $L_{n}$ is distributed in a wider range from $0.0$ to $2.0$, giving a distribution width of $2.0$, and the periodicity of $L_{n}$ disappears, as shown in Fig.\hspace{0.8mm}2(c) and Fig.\hspace{0.8mm}3(c).
A set of peaks extending from $n=0$ to $n=5018$ and beyond with no clear maxima
in the Fourier coefficients of consecutive point spacings
reflects the nonintegrability of the orbit.

\begin{figure}[htb]
 \begin{center}
  \begin{minipage}{62mm}
   \begin{center}
    \unitlength=2mm
     \special{epsfile=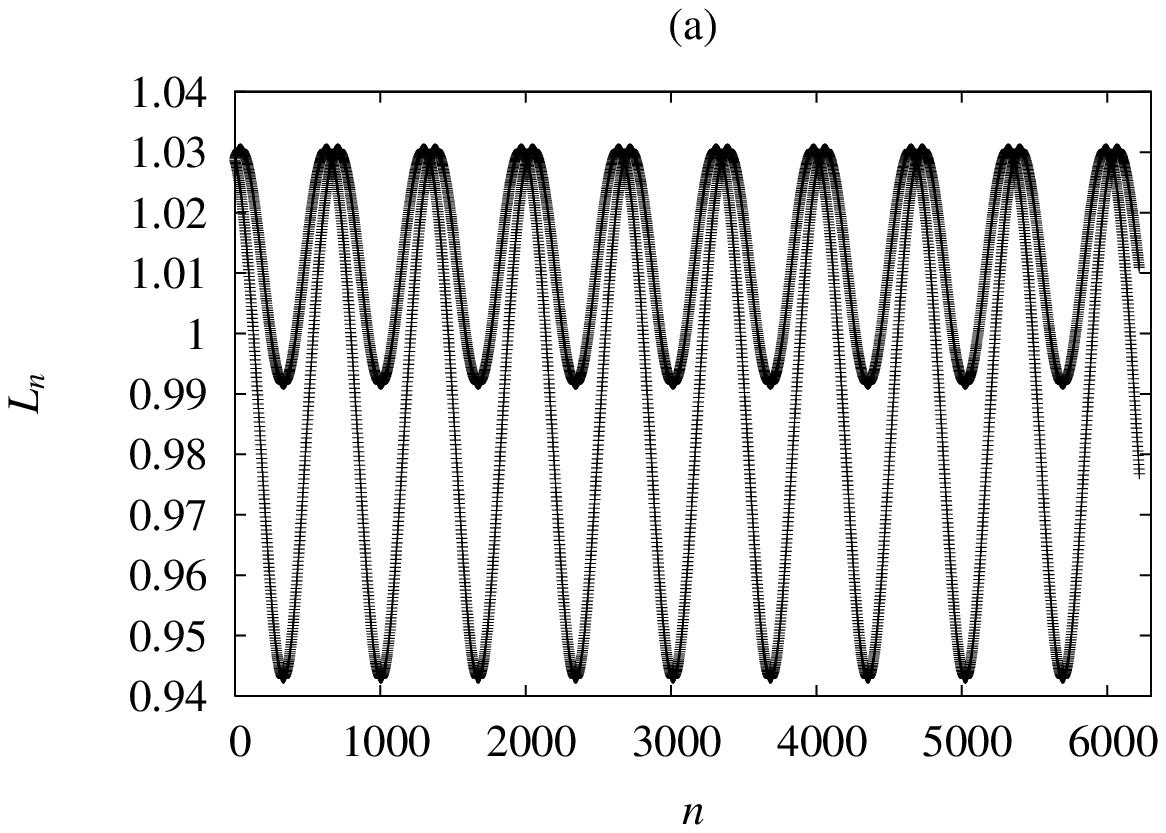 hscale=0.55 vscale=0.55}
      \vspace{52mm}
        \end{center}
         \end{minipage}
          \hspace{8mm}
         \begin{minipage}{62mm}
        \begin{center}
       \unitlength=2mm
      \special{epsfile=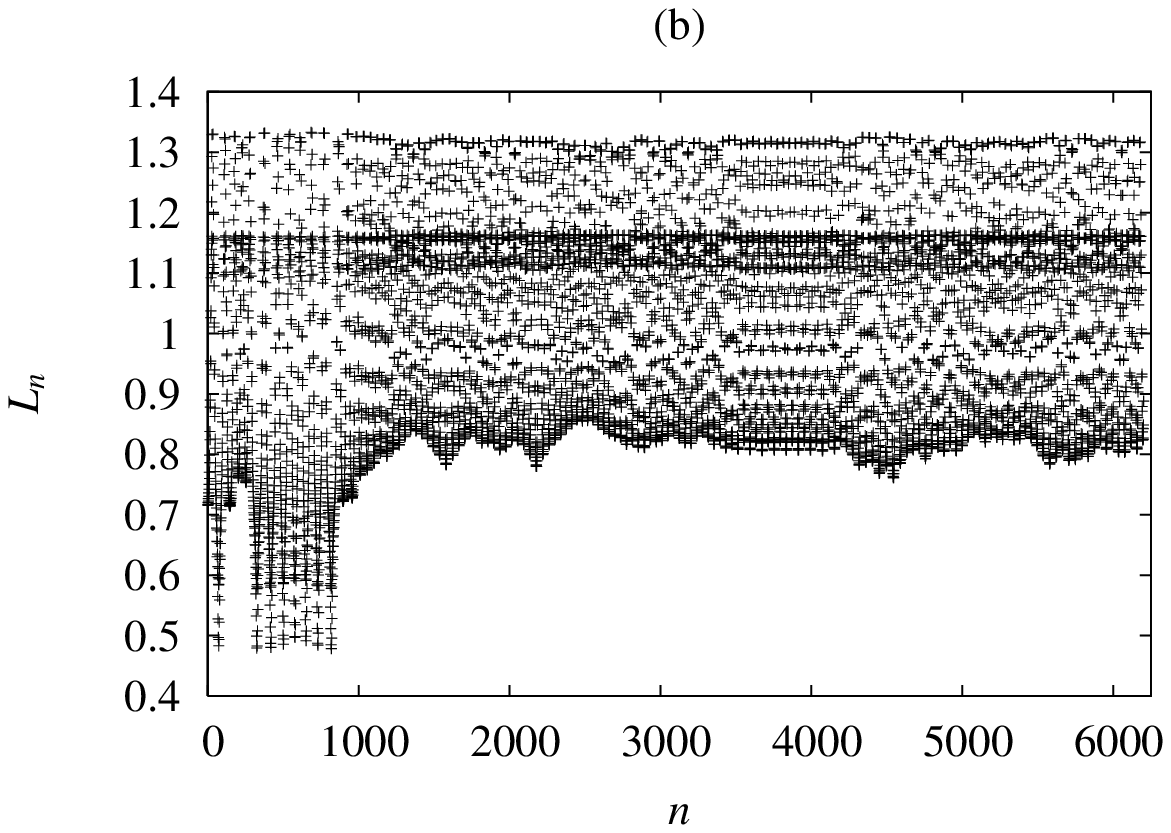 hscale=0.55 vscale=0.55}
     \vspace{52mm}
   \end{center}
  \end{minipage}
 \end{center}
 \begin{center}
  \begin{minipage}{62mm}
   \begin{center}
    \unitlength=2mm
     \special{epsfile=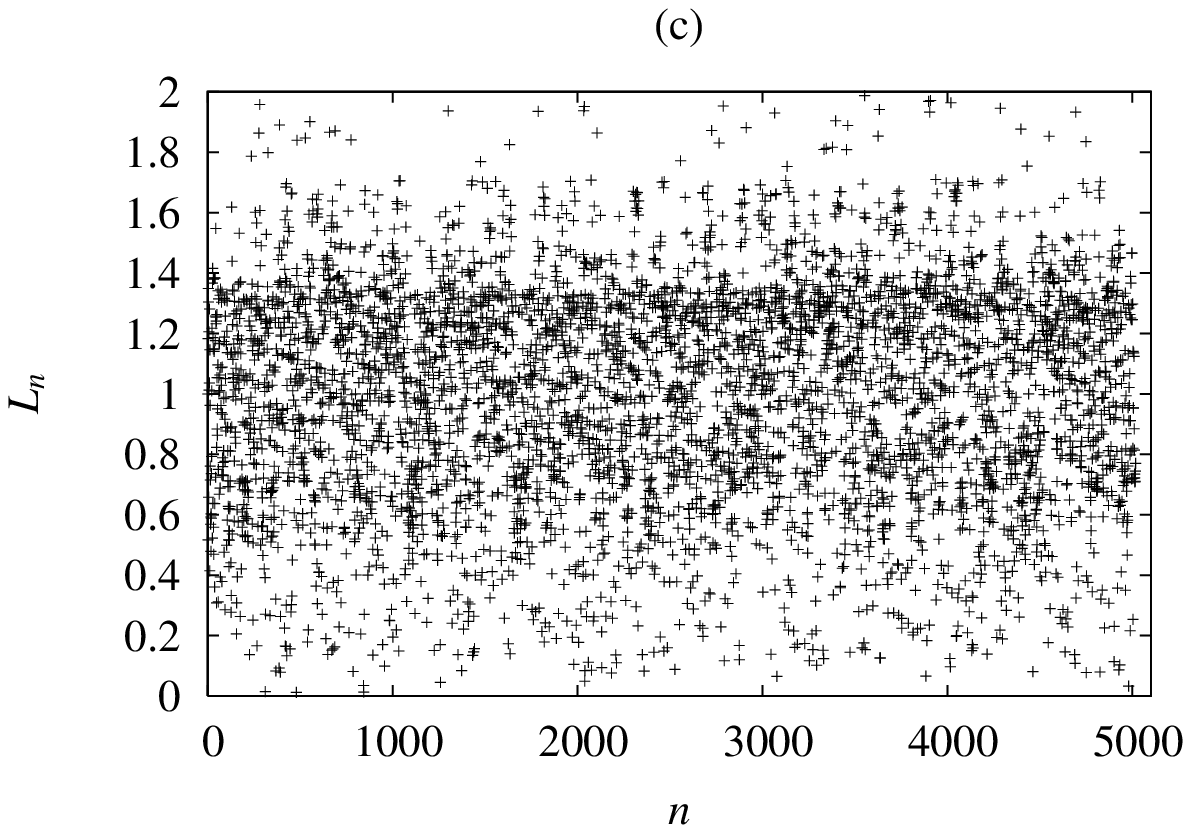 hscale=0.55 vscale=0.55}
      \vspace{53mm}
        \end{center}
         \end{minipage}
          \hspace{8mm}
         \begin{minipage}{62mm}
        \begin{center}
       \unitlength=2mm
      \special{epsfile=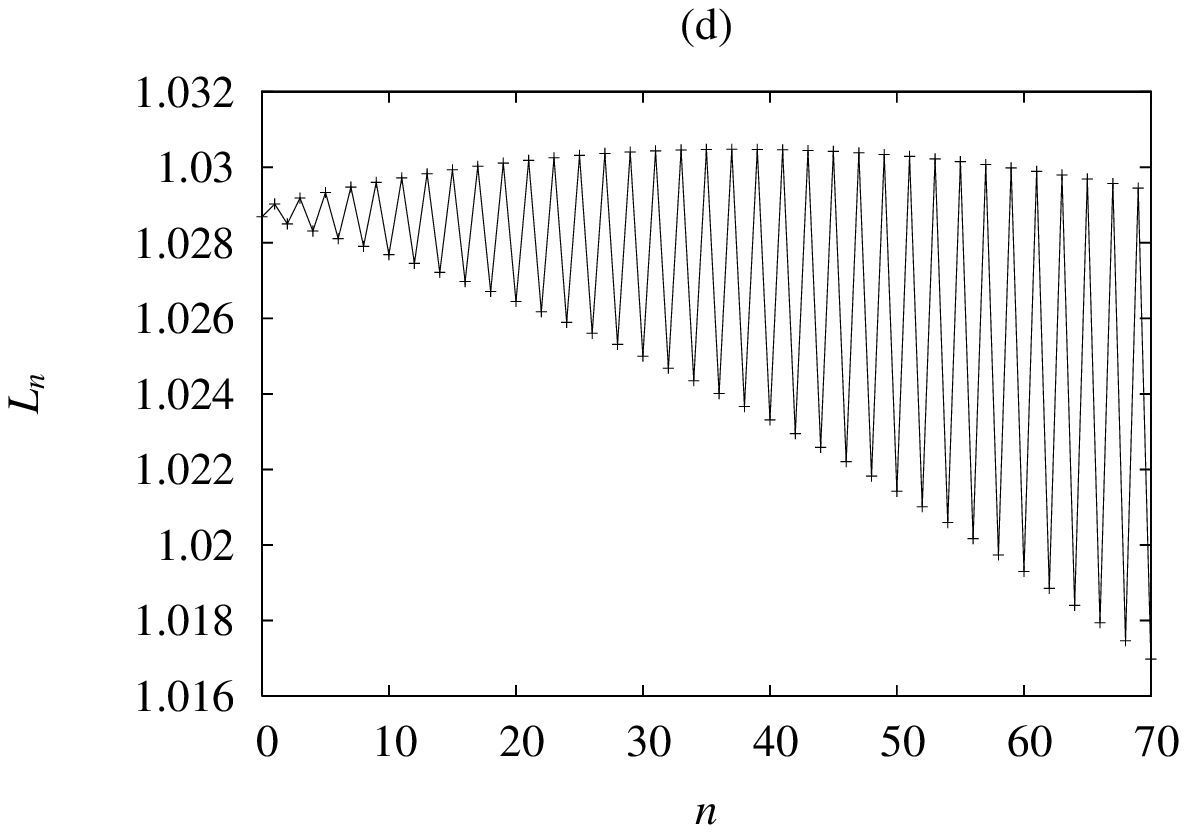 hscale=0.55 vscale=0.55}
     \vspace{53mm}
   \end{center}
  \end{minipage}
 \end{center}
       \caption{Consecutive phase-space point spacings computed using (a) $V=-0.009$, (b) $V=-0.045$, and (c) $V=-0.110$. The data set of Fig. 1(a)-(c) was used. (d) Short-time development of $L_{n}$ for $V=-0.009$.
}
\end{figure}

\begin{figure}[htb]
 \begin{center}
  \begin{minipage}{36mm}
   \begin{center}
    \unitlength=2mm
     \special{epsfile=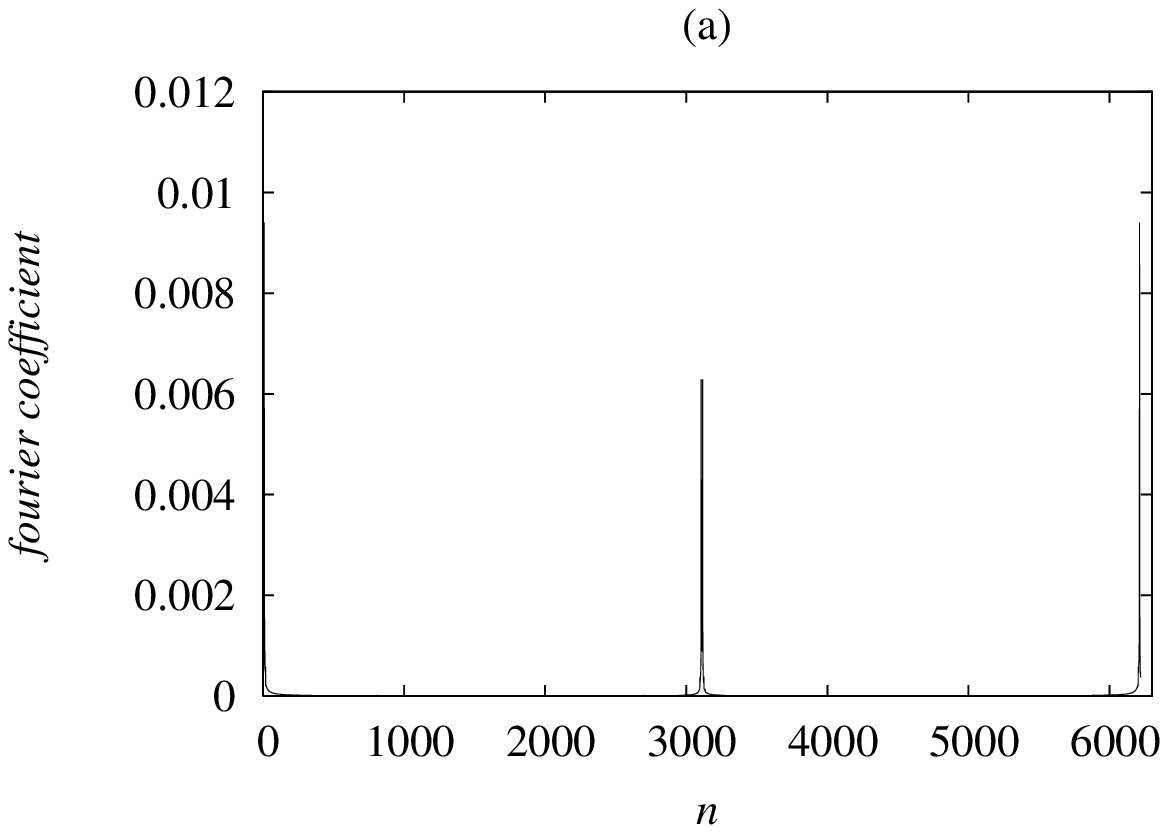 hscale=0.37 vscale=0.53}
      \vspace{51mm}
        \end{center}
         \end{minipage}
          \hspace{8mm}
         \begin{minipage}{36mm}
        \begin{center}
       \unitlength=2mm
      \special{epsfile=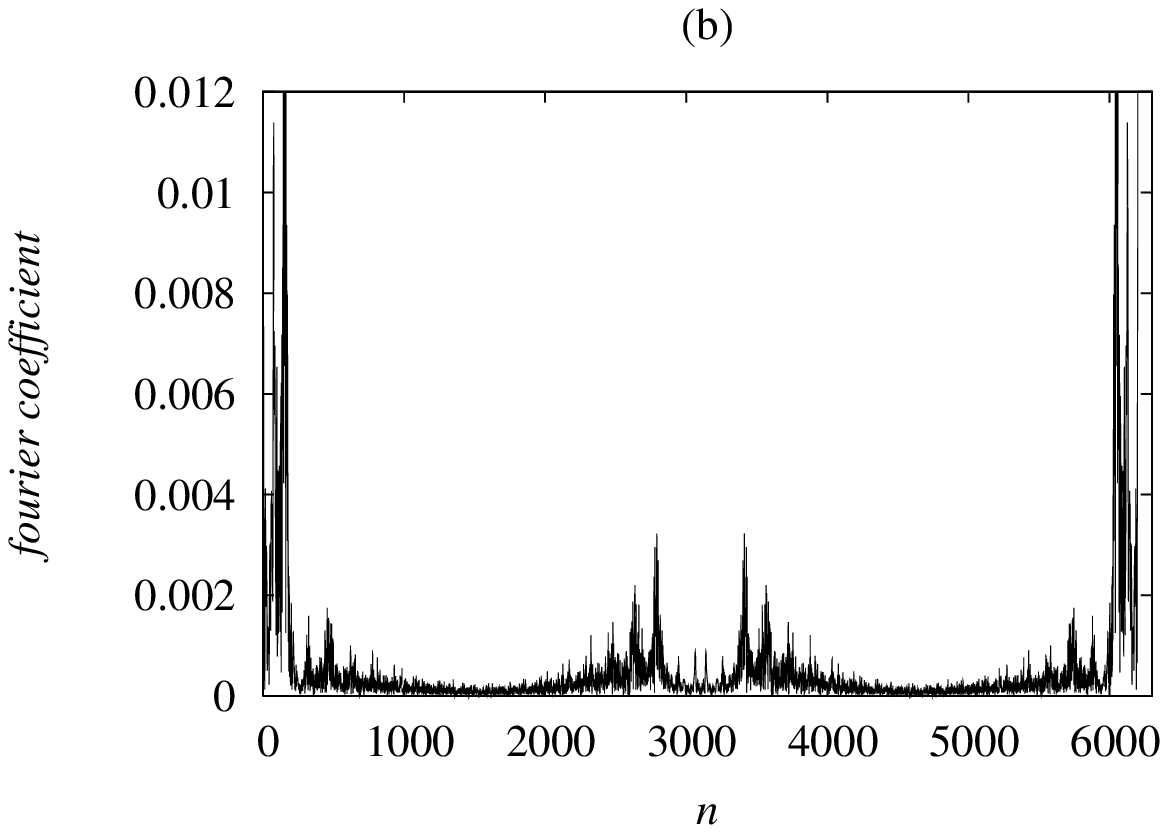 hscale=0.37 vscale=0.53}
     \vspace{51mm}
   \end{center}
  \end{minipage}
          \hspace{8mm}
         \begin{minipage}{36mm}
        \begin{center}
       \unitlength=2mm
      \special{epsfile=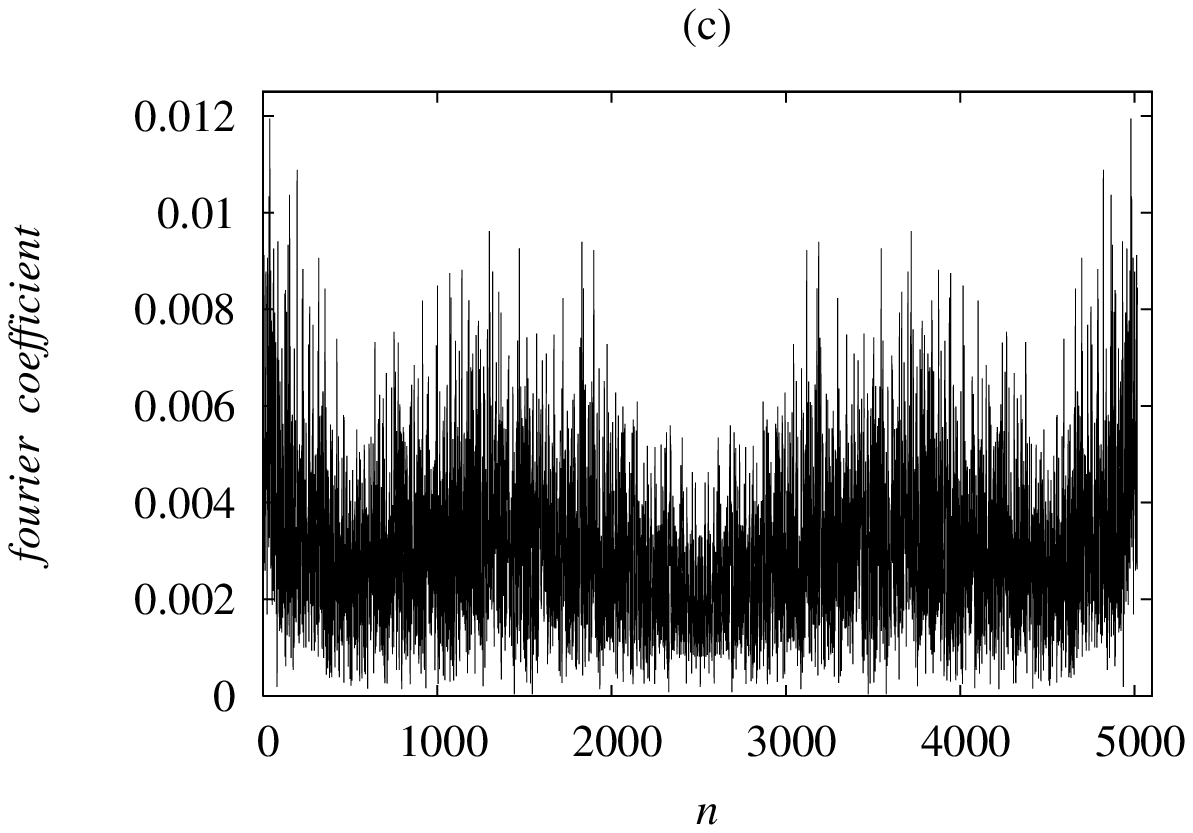 hscale=0.37 vscale=0.53}
     \vspace{51mm}
   \end{center}
  \end{minipage}
 \end{center}
       \caption{Root-mean-square of Fourier cosine coefficients and Fourier
sine coefficients for the consecutive phase-space point spacings depicted in
Fig.\hspace{0.8mm}2. (a) $V=-0.009$, (b) $V=-0.045$, and (c) $V=-0.110$.}
         \end{figure}

\section{Symmetry violation for consecutive phase-space point spacings and the relationship to chaos}

We will now examine a quantitative estimate of chaos that uses the absolute value of the correlation coefficient for the rates of elongation and contraction of $L_{n}$.
In the case where $\frac{L_{n +1}}{L_{n}}>1$, the rate of elongation is defined as $R_{m}\equiv \frac{L_{n +1}}{L_{n}}$, and
in the case where $\frac{L_{n +1}}{L_{n}}<1$, the rate of contraction is defined as $r_{m}\equiv \frac{L_{n +1}}{L_{n}}$.
For example, the values $\frac{L_{2}}{L_{1}}=1.1$, $\frac{L_{3}}{L_{2}}=1.2$, $\frac{L_{4}}{L_{3}}=0.9$, $\frac{L_{5}}{L_{4}}=0.8$, $\frac{L_{6}}{L_{5}}=1.3$, and $\frac{L_{7}}{L_{6}}=0.7$ give rise to elongation rates $R_{1}=1.1$, $R_{2}=1.2$, and $R_{3}=1.3$ and  contraction rates $r_{1}=0.9$, $r_{2}=0.8$, and $r_{3}=0.7$.
The average rates of elongation and contraction for a number $M$ of phase-space point spacings
are given by
\begin{eqnarray}
\bar{R}=\frac{1}{M}\sum_{m =1}^{M}R_{m}\hspace{1mm},\hspace{1mm}
\bar{r}=\frac{1}{M}\sum_{m =1}^{M}r_{m} .
\end{eqnarray}
One may introduce a correlation coefficient\cite{rf:6} for $R_{m}$ and
$r_{m}$ that takes the form
\begin{eqnarray}
C_{R_{m}r_{m}}=\frac{\displaystyle\sum_{m =1}^{M}(R_{m}-\bar{R})(r_{m}-\bar{r})}
{\sqrt{\displaystyle\sum_{m =1}^{M}(R_{m}-\bar{R})^{2}}
\sqrt{\displaystyle\sum_{m =1}^{M}(r_{m}-\bar{r})^{2}}}\hspace{0.8mm}.
\end{eqnarray}
Now, let us begin our analysis by considering the regular-to-chaos transition visible in Figs.\hspace{0.8mm}1(a)-(c) and Figs.\hspace{0.8mm}2(a)-(c).
Since the consecutive phase-space point spacings are periodic in the case of a weak interaction, the time dependences of $R_{m}$ and $r_{m}$
are symmetric with respect to $\frac{L_{n +1}}{L_{n}}=1$, as shown
in Fig.\hspace{0.8mm}4(a).
In this case, the absolute value of the correlation coefficient is close to 1.0.
In Fig.\hspace{0.8mm}4, note that the time series of $R_{m}$ and $r_{m}$ are independent of each other.
As the interaction strength increases only slightly, the orbit appears as shown in Fig.\hspace{0.8mm}1(b), and $R_{m}$ and $r_{m}$ now behave as shown in Fig.\hspace{0.8mm}4(b). 
This orbit is, however, not integrable and it can be seen that the symmetry between $R_{m}$ and $r_{m}$ is approximately maintained as shown in Figs.\hspace{0.8mm}4(b) and 4(d). 
When the interaction strength increases as shown in Fig.\hspace{0.8mm}1(c), one may not expect symmetric behavior between $R_{m}$ and $r_{m}$, and this is demonstrated in Fig.\hspace{0.8mm}4(c).
In this case, as seen from Fig.\hspace{0.8mm}5, the absolute values of the correlation coefficients are close to zero.

\begin{figure}[htb]
 \begin{center}
  \begin{minipage}{62mm}
   \begin{center}
    \unitlength=2mm
     \special{epsfile=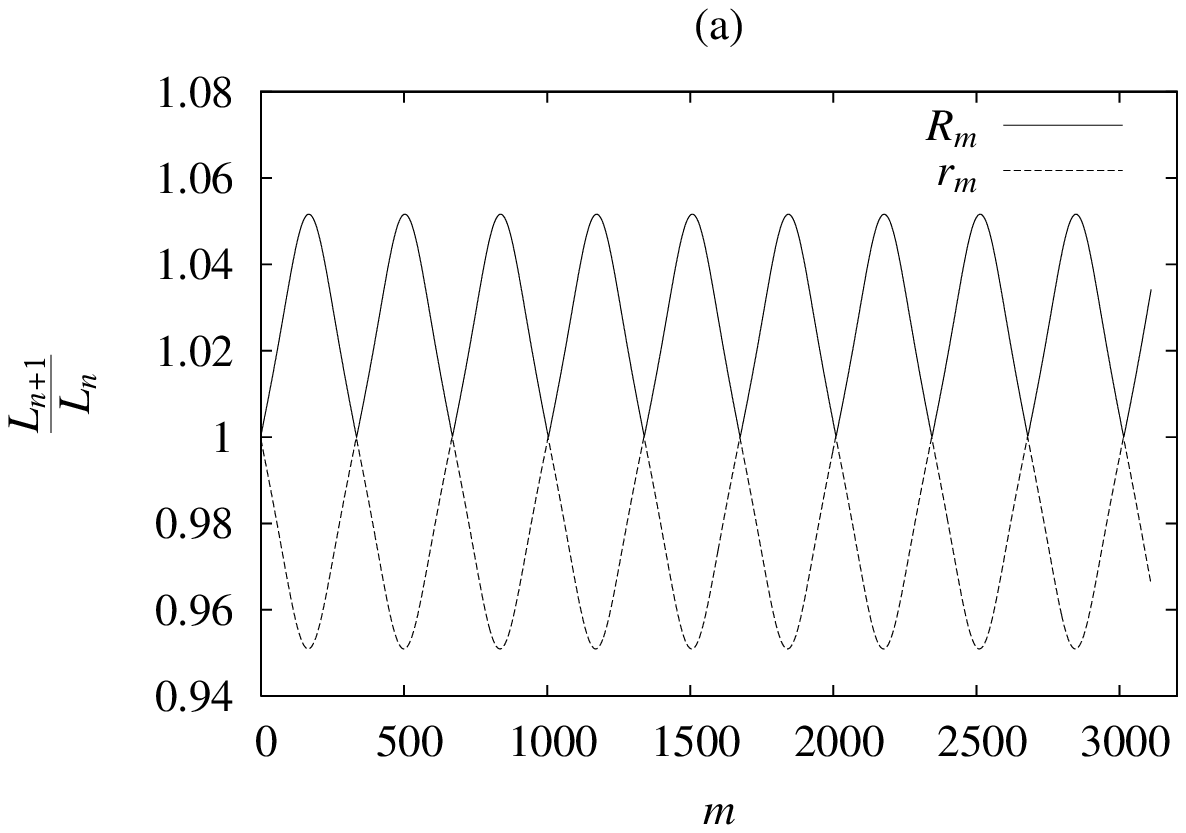 hscale=0.55 vscale=0.55}
      \vspace{52mm}
        \end{center}
         \end{minipage}
          \hspace{8mm}
         \begin{minipage}{62mm}
        \begin{center}
       \unitlength=2mm
      \special{epsfile=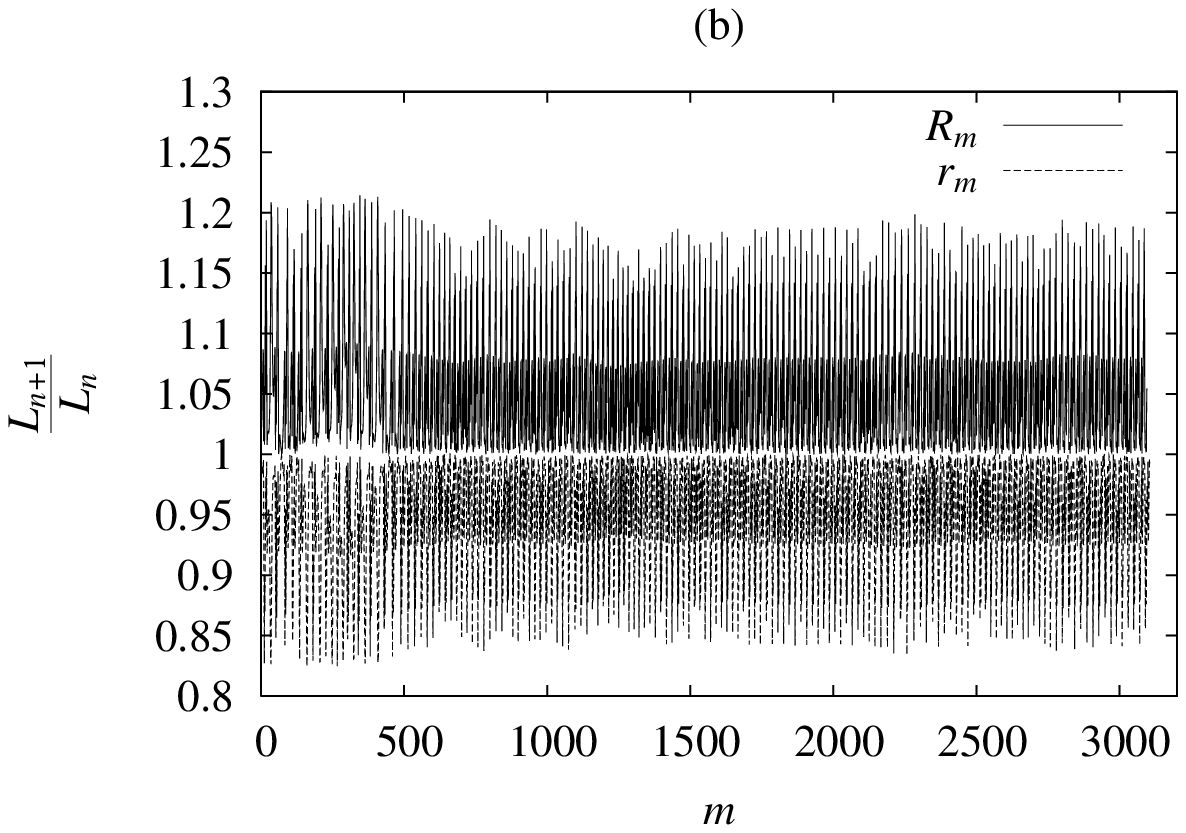 hscale=0.55 vscale=0.55}
     \vspace{52mm}
   \end{center}
  \end{minipage}
 \end{center}
 \begin{center}
  \begin{minipage}{62mm}
   \begin{center}
    \unitlength=2mm
     \special{epsfile=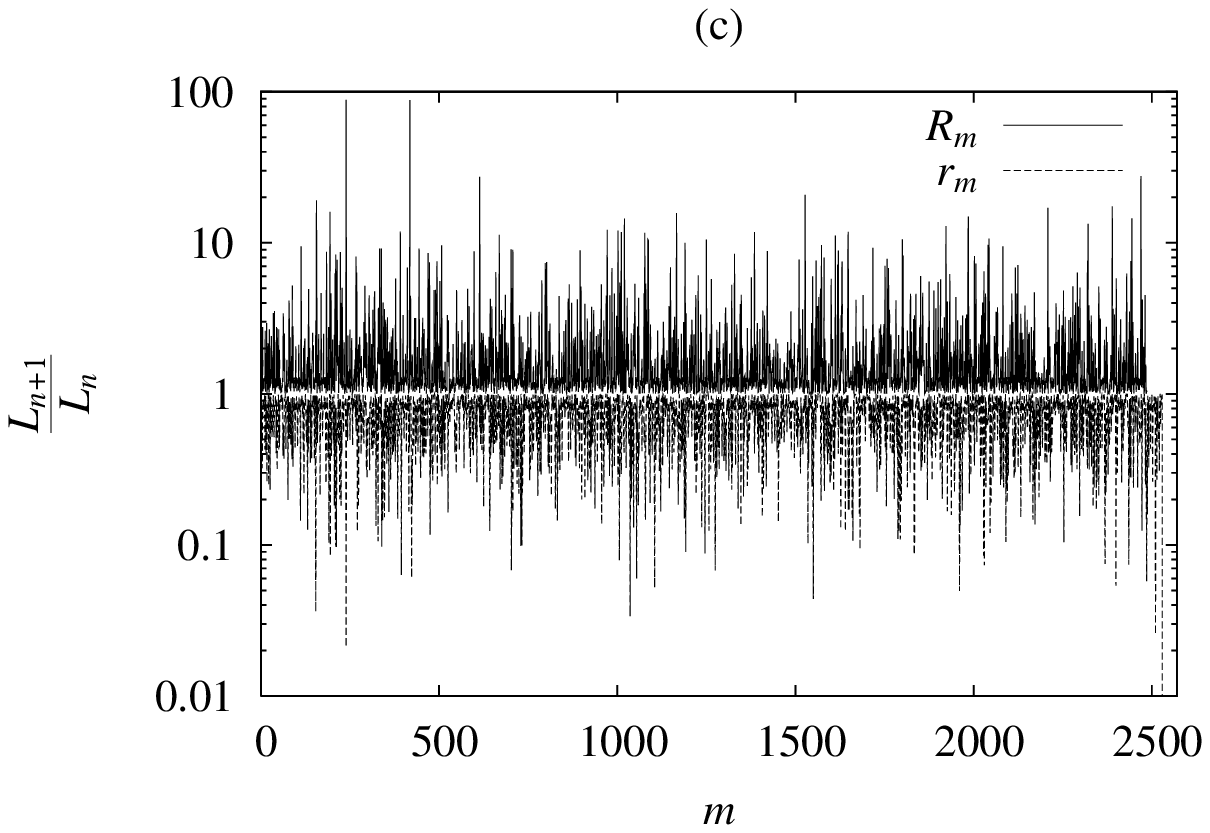 hscale=0.55 vscale=0.55}
      \vspace{53mm}
        \end{center}
         \end{minipage}
          \hspace{8mm}
         \begin{minipage}{62mm}
        \begin{center}
       \unitlength=2mm
      \special{epsfile=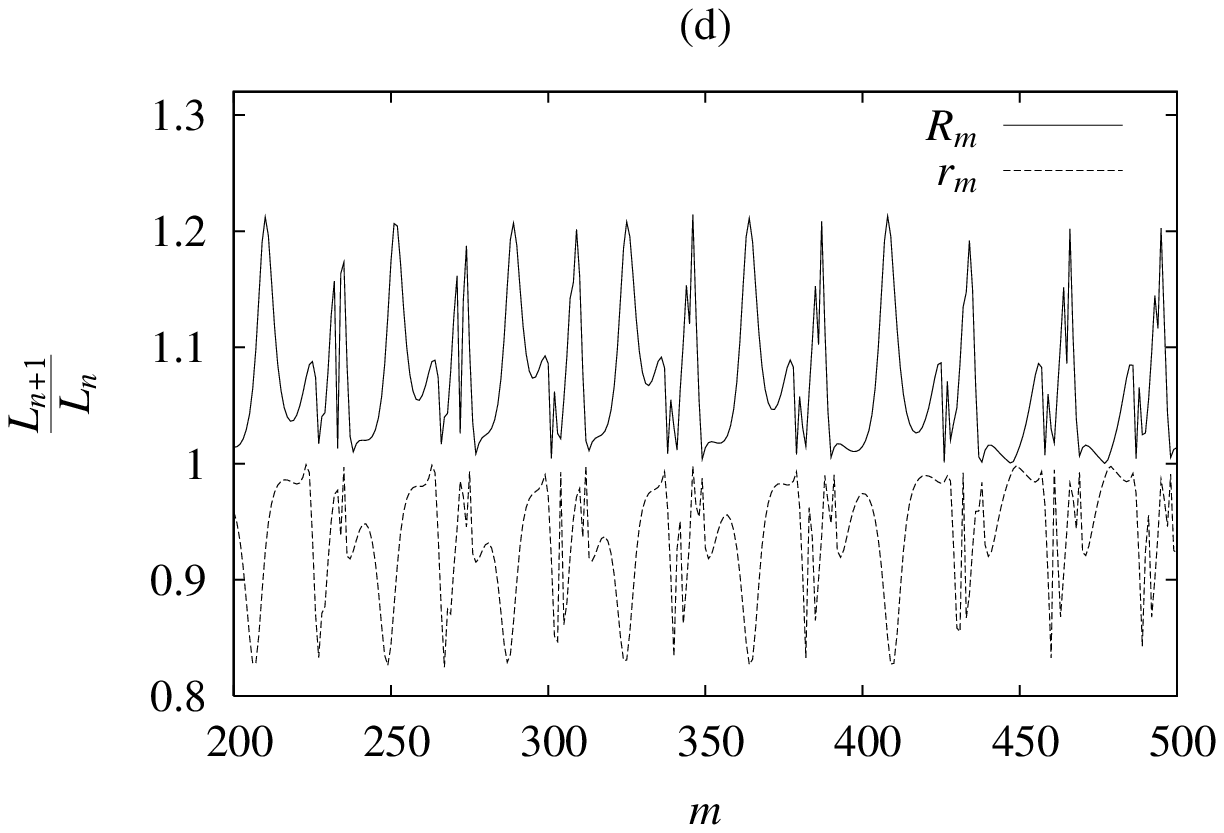 hscale=0.55 vscale=0.55}
     \vspace{53mm}
   \end{center}
  \end{minipage}
 \end{center}
       \caption{Elongation rate and contraction rate of phase-space point spacings. 
(a) $V=-0.009$, (b) $V=-0.045$, and (c) $V=-0.110$. (d) Short-time development of $\frac{L_{n+1}}{L_{n}}$ for $V=-0.045$.
}
\end{figure}

\clearpage

\begin{figure}[ht]
 \begin{minipage}{66mm}
   \begin{center}
     \hspace{1mm}\special{epsfile=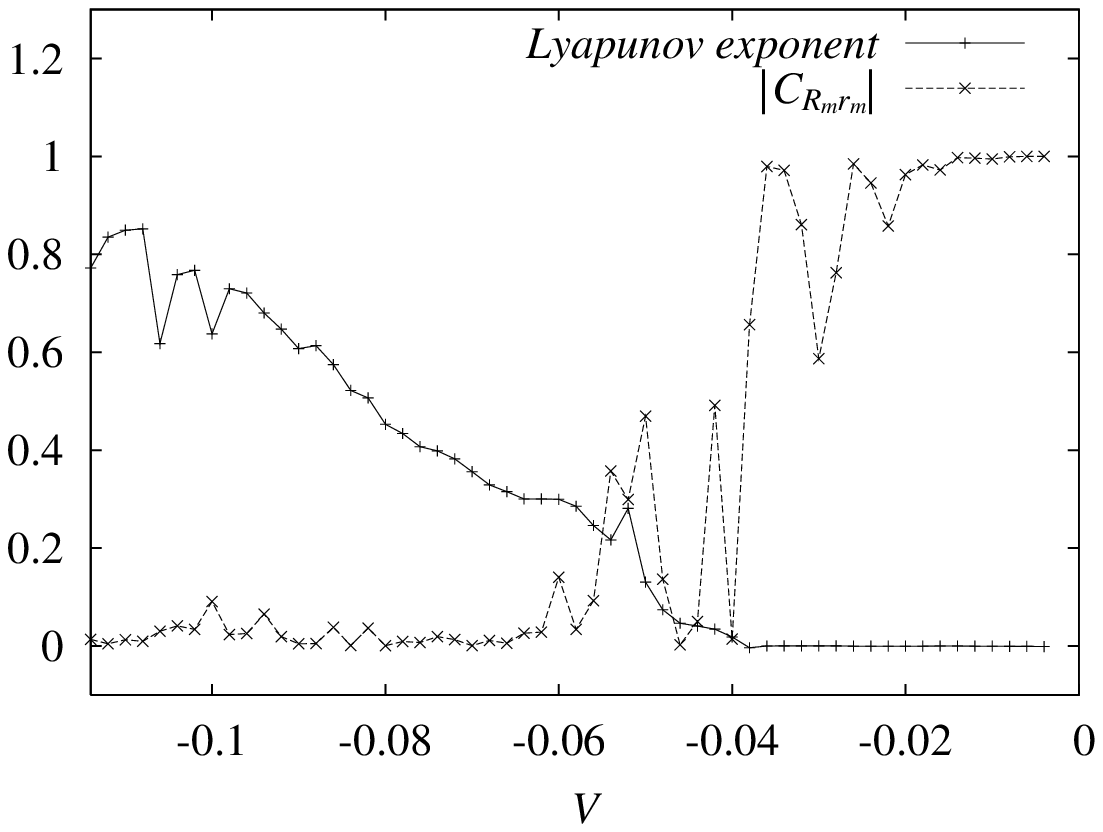 hscale=0.7 vscale=0.6}
     \vspace{56mm}
  \end{center}
  \end{minipage}
       \caption{Graph of both the Lyapunov exponent and the absolute value of the correlation coefficient in the transition process from integrability to nonintegrability. Details of the comparison are described in section 4.}
         \end{figure}

\section{Features of the correlation coefficient analysis}

In this section, the process is analyzed using the absolute value of the correlation
coefficient and the Lyapunov exponent, for the case in which the periodic orbit transitions to the nonintegrable orbit. 

In the case with weak interaction, the absolute value of the correlation coefficient quickly reaches $1.0$ as shown in Fig.\hspace{0.8mm}6(a). 
When the interaction strength increases, the absolute values of the correlation coefficients are close to $0.0$, as shown in the plot for $V=-0.110$ in Fig.\hspace{0.8mm}6(a).

\begin{figure}[htb]
 \begin{center}
  \begin{minipage}{64mm}
   \begin{center}
    \unitlength=2mm
     \special{epsfile=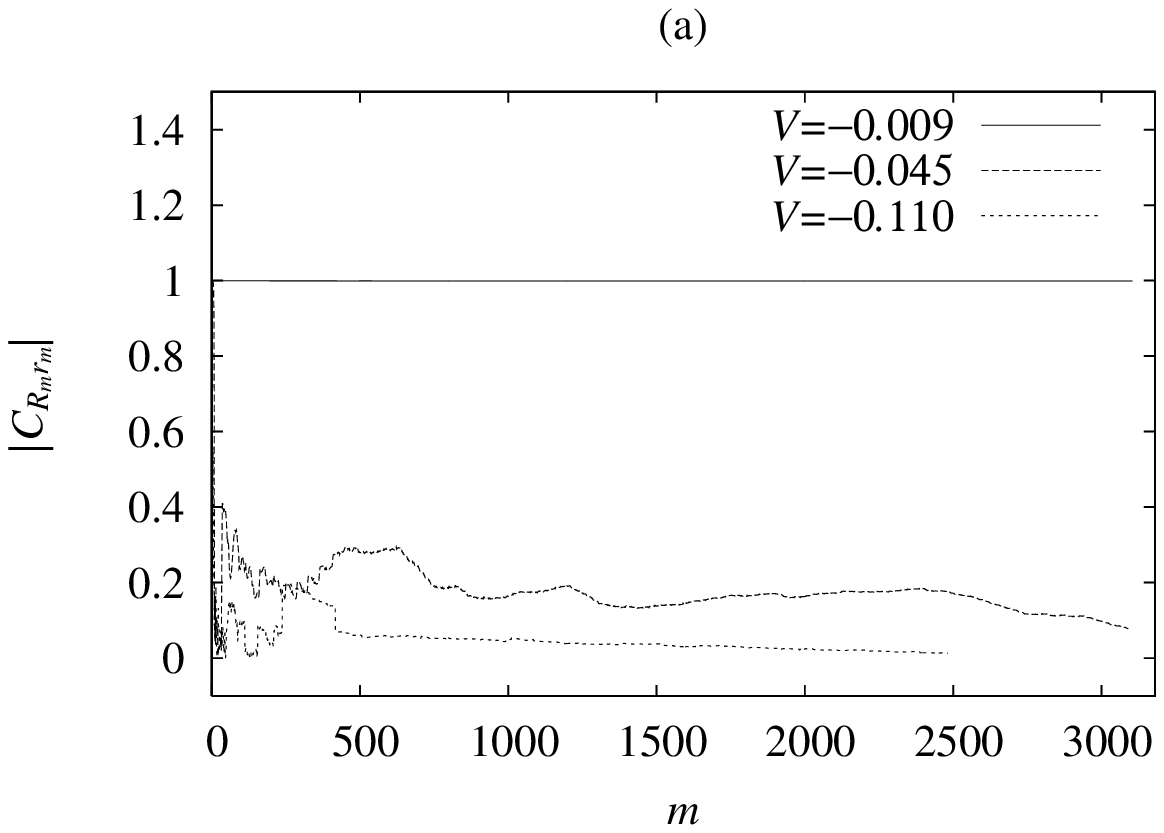 hscale=0.55 vscale=0.55}
      \vspace{53mm}
        \end{center}
         \end{minipage}
          \hspace{5mm}
         \begin{minipage}{64mm}
        \begin{center}
       \unitlength=2mm
      \special{epsfile=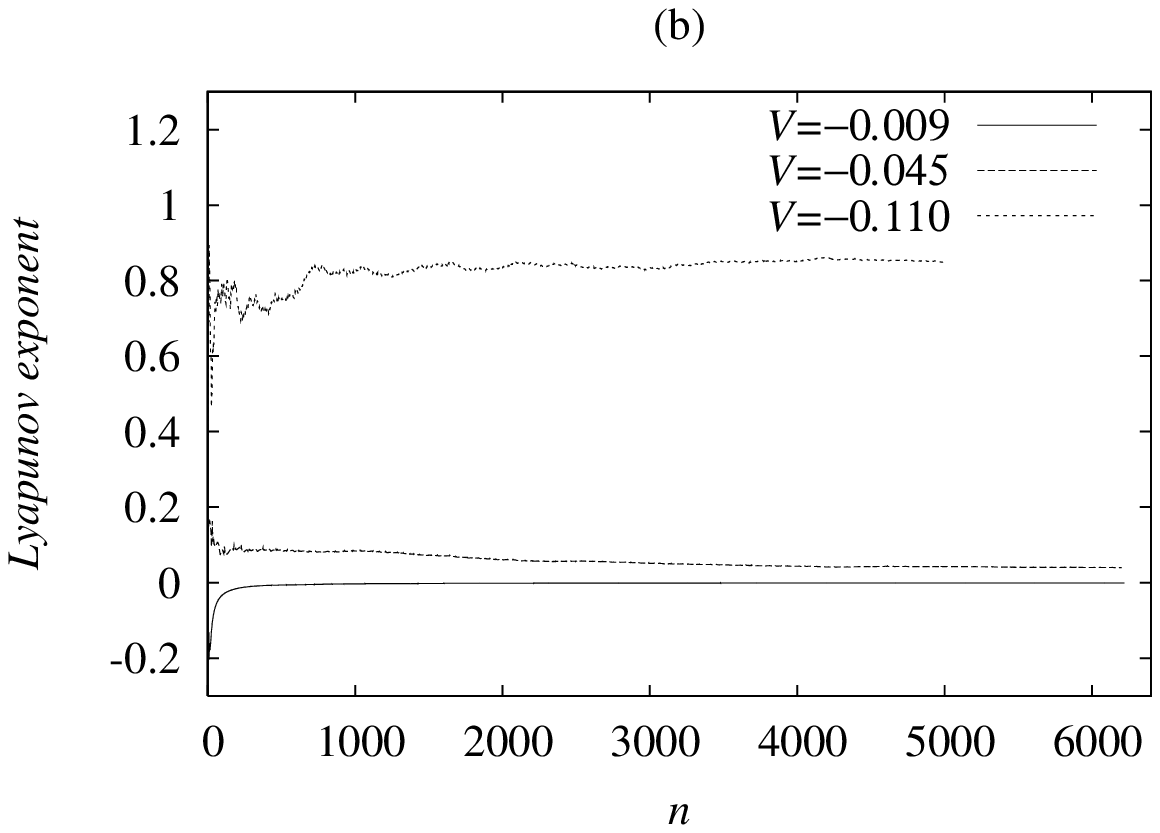 hscale=0.55 vscale=0.55}
     \vspace{53mm}
   \end{center}
  \end{minipage}
       \caption{(a) Time dependence of the absolute value of the correlation coefficient for Figs. 1(a)-(c).
(b) Time dependence of the Lyapunov exponent for Figs. 1(a)-(c).
The number of data points in (a) is roughly half that in (b) due to the division into elongation and contraction rates used to calculate the correlation coefficient.}
 \end{center}
\end{figure}

When the orbit is completely integrable, the absolute value of the correlation coefficient has exactly $1.0$ as an upper bound. 
On the other hand, if the orbit is nonintegrable, it has $|C_{R_{m}r_{m}}|=0.0$ as a lower bound. 
The Lyapunov exponent takes negative values when the system is integrable and positive values
when the system is non-integrable. 
However, the upper and lower limits of the value of the Lyapunov exponent are not clear, as depicted in Fig. 6(b).

\section{Asymmetric behavior in consecutive phase-space point spacings and nonintegrability in the H\'{e}non-Heiles system}

To date, the analysis of $|C_{R_{m}r_{m}}|$ has been performed using the specific {\textit {SU}}(3) Hamiltonian with fixed energy $E$=40. 
To confirm the usefulness of our analysis using $|C_{R_{m}r_{m}}|$, let us apply our method to the H\'{e}non-Heiles system\cite{rf:1} and also study what happens when the energy of the system is varied. 

The Hamiltonian is given by
\begin{eqnarray}
H(q^{(1)},p^{(1)};q^{(2)},p^{(2)})=\frac{1}{2}({p^{(1)}}^{2}+{p^{(2)}}^{2})+\frac{1}{2}({q^{(1)}}^{2}+{q^{(2)}}^{2})+{q^{(1)}}^{2}q^{(2)}-\frac{1}{3}{q^{(2)}}^{3}\hspace{1mm}.\nonumber\\
\end{eqnarray}

If the energy $E$ is sufficiently small, 
the Poincar$\acute{\textrm{e}}$ section map turns out to be integrable, as shown in Fig. 7(a). 
In this case, it is found that a degree of symmetry occurs between the $R_{m}$ and $r_{m}$ as shown in Fig. 8(a). Since this symmetry occurs, the absolute value of the correlation coefficient becomes $0.5091691$ shown in Fig. 9(a). 
In Fig.\hspace{0.8mm}7(b), 
the successive intersections of the trajectory with the 
Poincar$\acute{\textrm{e}}$ section plane lie on twenteen closed curves with ten hyperbolic fixed points located at the joins of adjacent closed curves. 
Nonintegrability appears in the H\'{e}non-Heiles system as energy increases. 
An example of this at $E=0.1669$ is shown in Fig. 7(c). 
From Figs. 8(c) and 9(a), it can be seen that the symmetry between $R_{m}$ and $r_{m}$ is violated with the result that the absolute value of the correlation coefficient approaches zero. 

\begin{figure}[htb]
 \begin{center}
  \begin{minipage}{36mm}
   \begin{center}
    \unitlength=2mm
     \special{epsfile=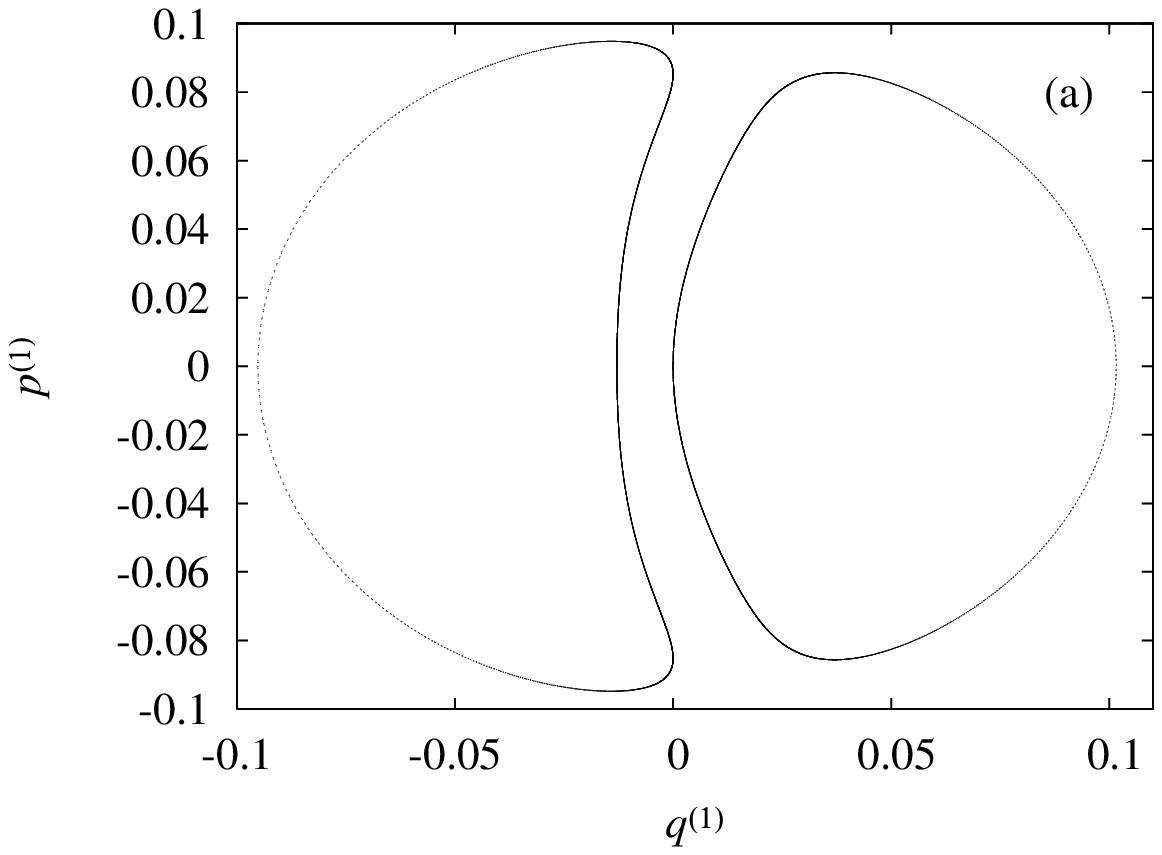 hscale=0.37 vscale=0.53}
      \vspace{51mm}
        \end{center}
         \end{minipage}
          \hspace{8mm}
         \begin{minipage}{36mm}
        \begin{center}
       \unitlength=2mm
      \special{epsfile=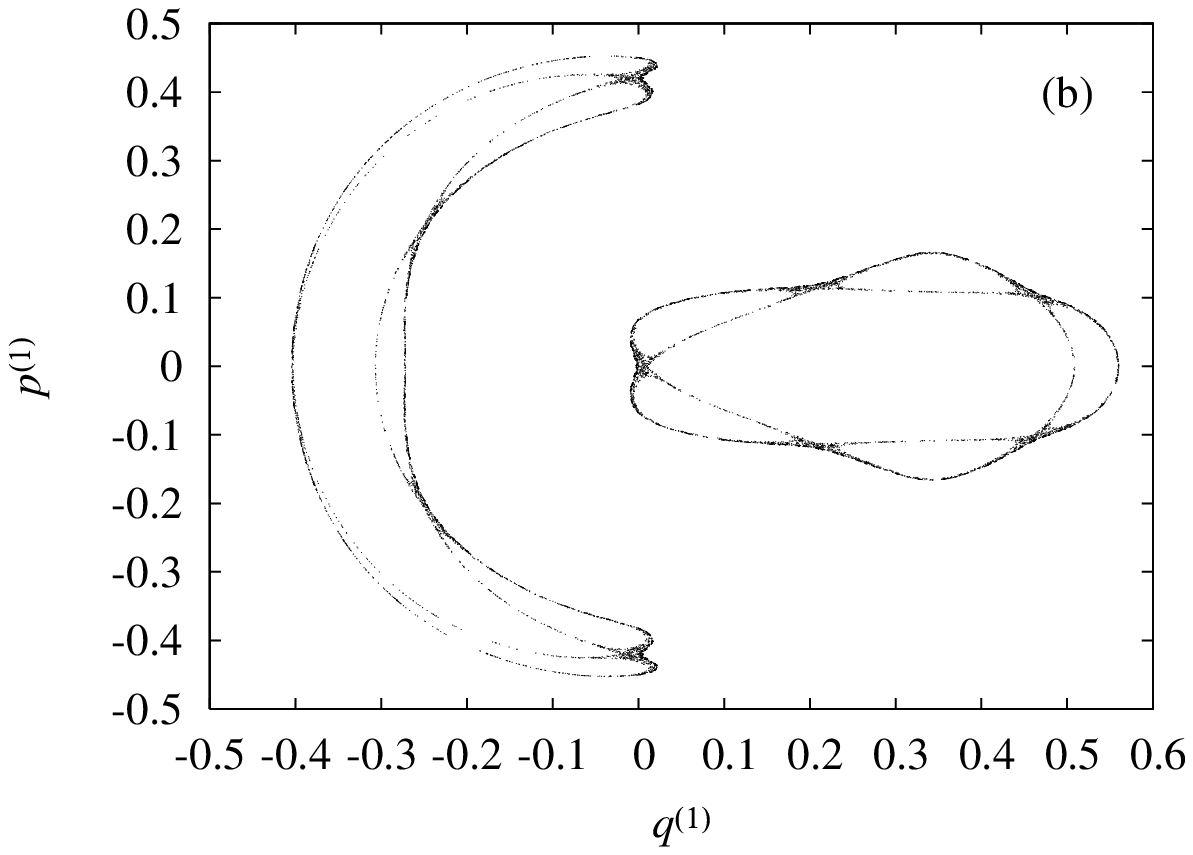 hscale=0.37 vscale=0.53}
     \vspace{51mm}
   \end{center}
  \end{minipage}
          \hspace{8mm}
         \begin{minipage}{36mm}
        \begin{center}
       \unitlength=2mm
      \special{epsfile=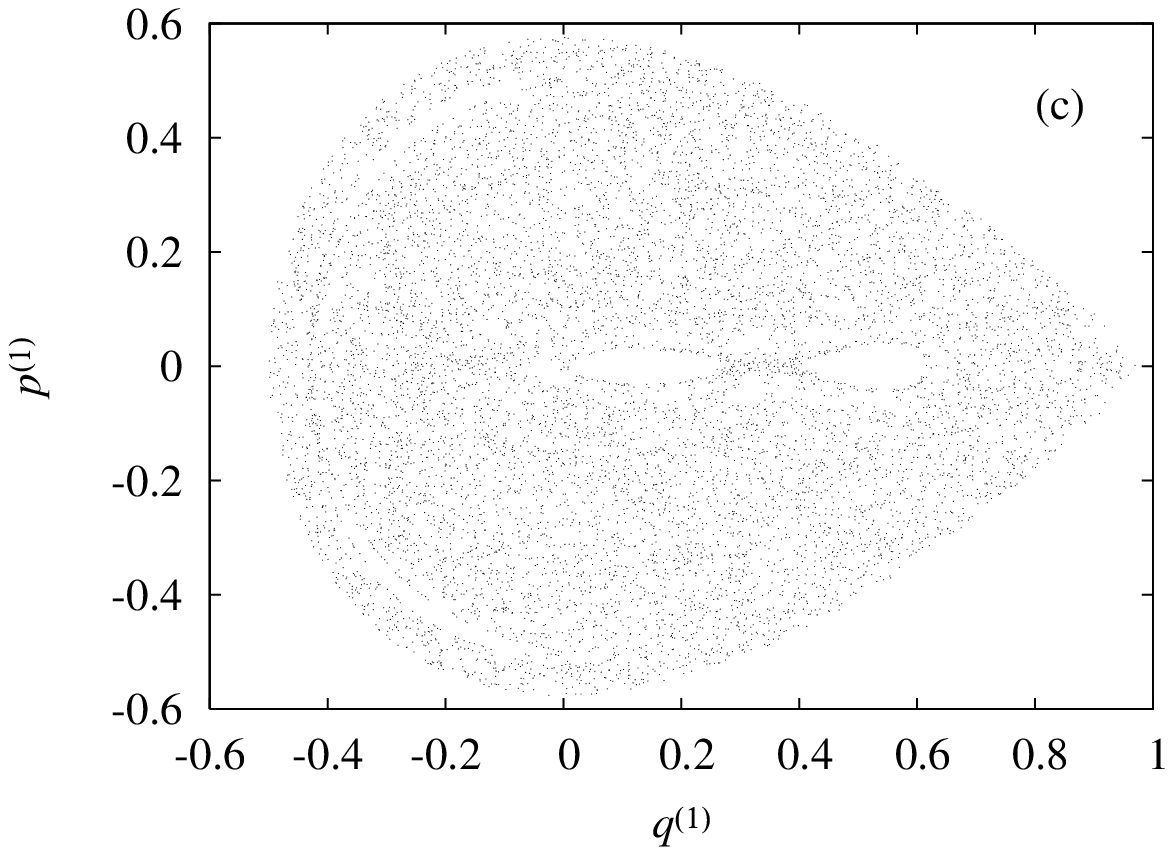 hscale=0.37 vscale=0.53}
     \vspace{51mm}
   \end{center}
  \end{minipage}
 \end{center}
       \caption{Surfaces of sections for the H\'{e}non-Heiles Hamiltonian with (a) $E=0.005$, (b) $E=0.11796899$, and (c) $E=0.1669$ from numerical integration.}
         \end{figure}

\begin{figure}[htb]
 \begin{center}
  \begin{minipage}{62mm}
   \begin{center}
    \unitlength=2mm
     \special{epsfile=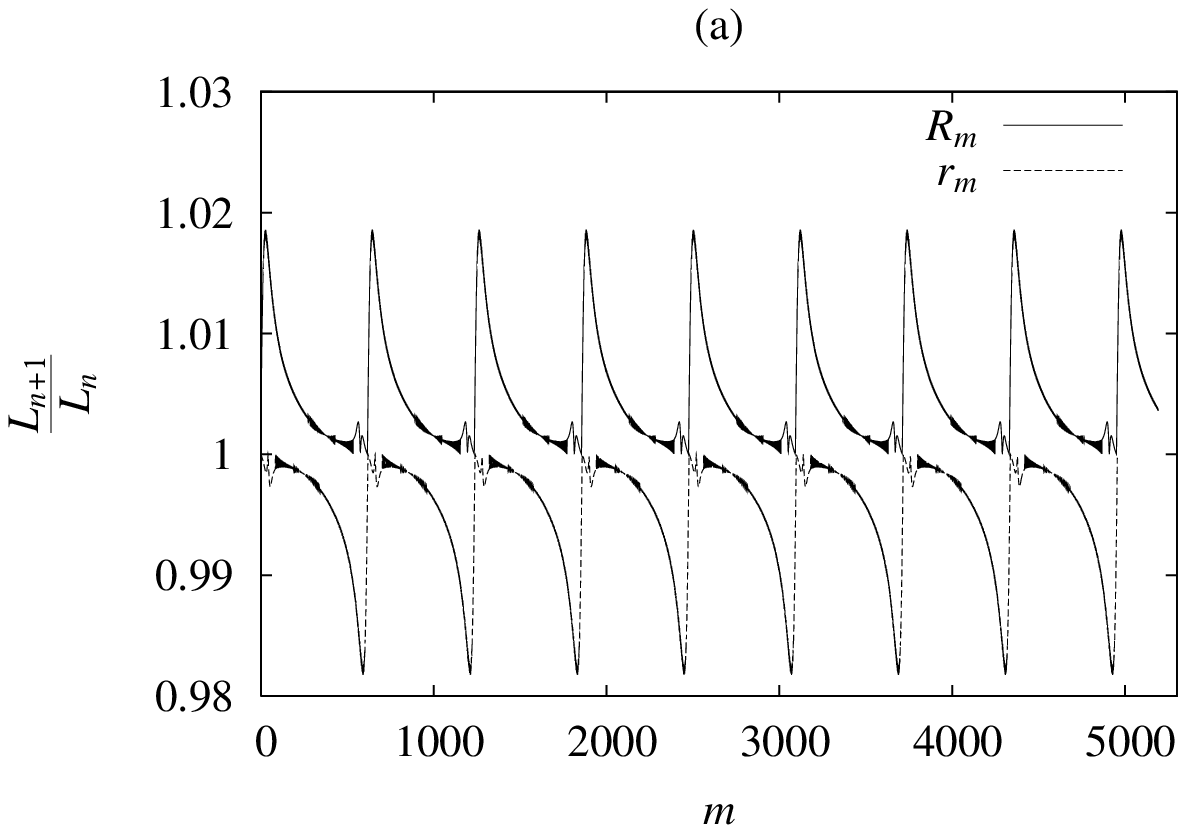 hscale=0.55 vscale=0.55}
      \vspace{52mm}
        \end{center}
         \end{minipage}
          \hspace{8mm}
         \begin{minipage}{62mm}
        \begin{center}
       \unitlength=2mm
      \special{epsfile=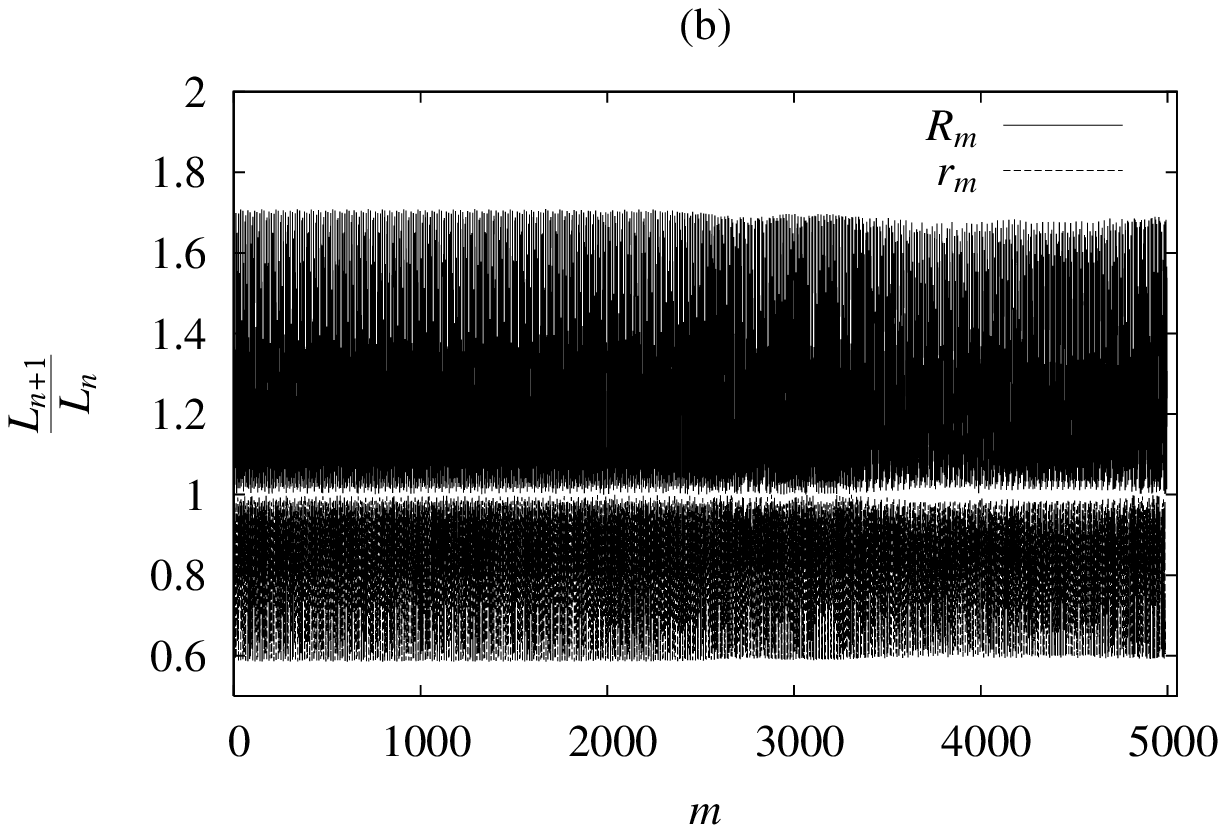 hscale=0.55 vscale=0.55}
     \vspace{52mm}
   \end{center}
  \end{minipage}
 \end{center}
 \begin{center}
  \begin{minipage}{62mm}
   \begin{center}
    \unitlength=2mm
     \special{epsfile=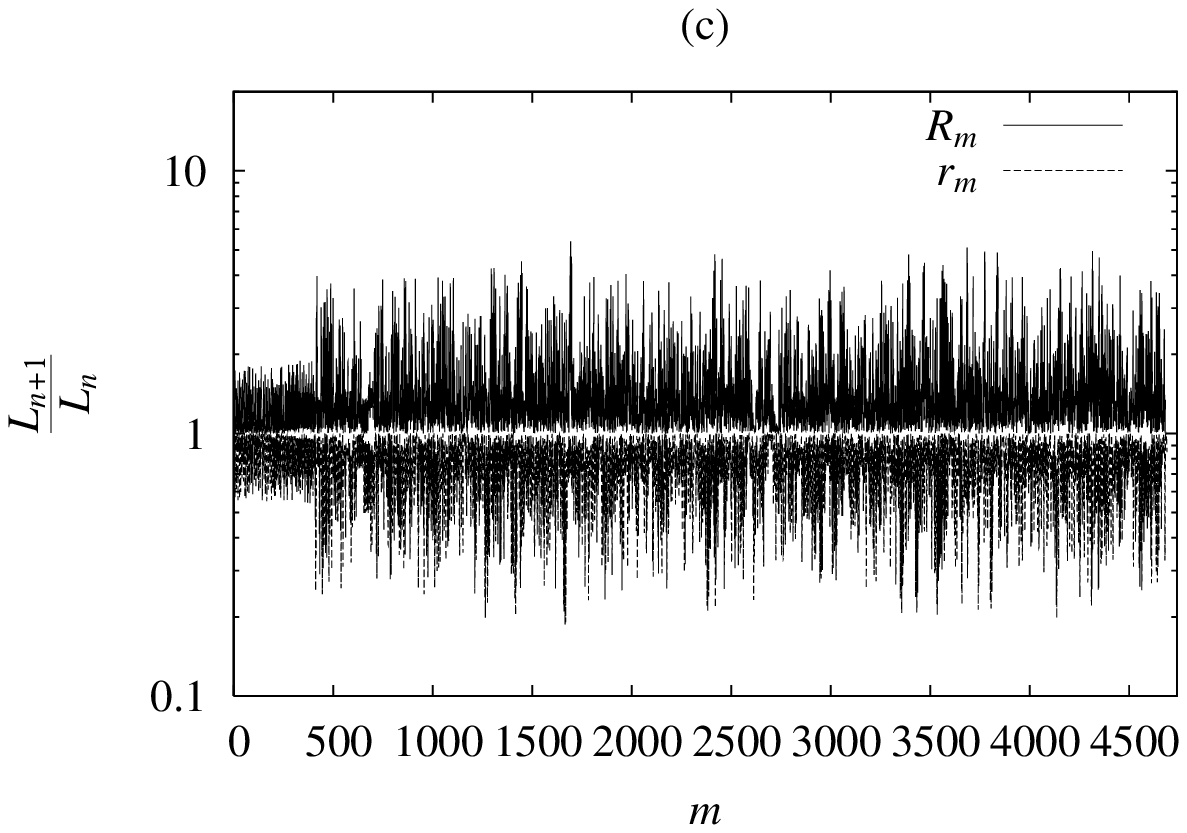 hscale=0.55 vscale=0.55}
      \vspace{53mm}
        \end{center}
         \end{minipage}
          \hspace{8mm}
         \begin{minipage}{62mm}
        \begin{center}
       \unitlength=2mm
      \special{epsfile=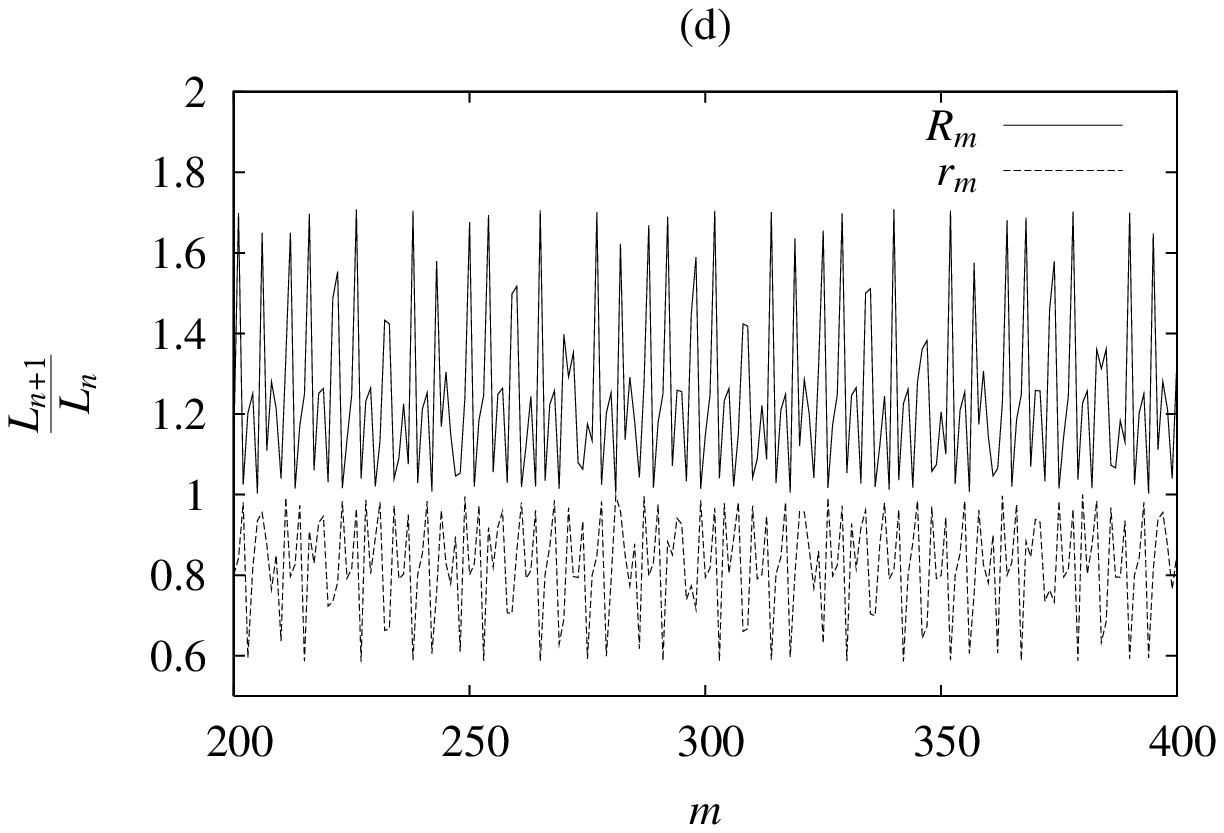 hscale=0.55 vscale=0.55}
     \vspace{53mm}
   \end{center}
  \end{minipage}
 \end{center}
       \caption{Elongation rate and contraction rate of phase-space point spacings. 
(a) $E=0.005$, (b) $E=0.11796899$, and (c) $E=0.1669$. 
(d) Short-time development of $\frac{L_{n+1}}{L_{n}}$ for $E=0.11796899$.
}
\end{figure}

\begin{figure}[htb]
 \begin{center}
  \begin{minipage}{64mm}
   \begin{center}
    \unitlength=2mm
     \special{epsfile=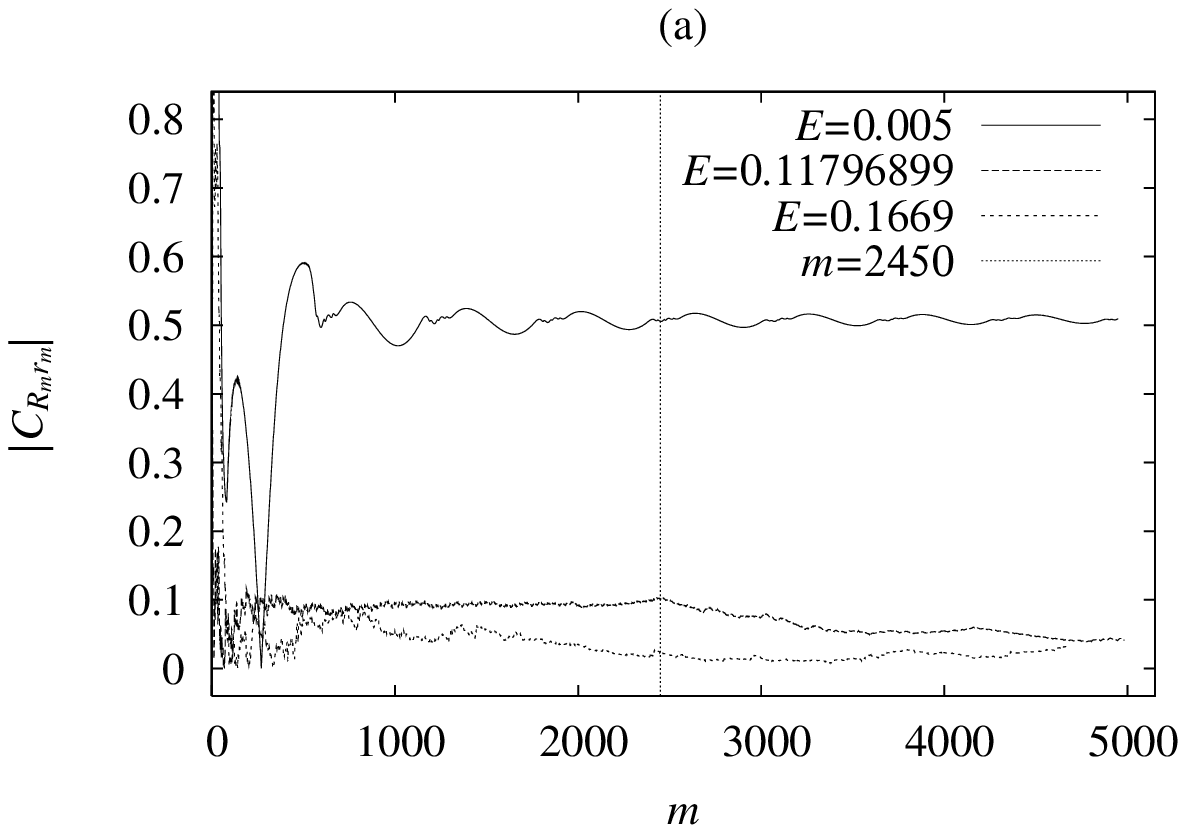 hscale=0.55 vscale=0.56}
      \vspace{53mm}
        \end{center}
         \end{minipage}
          \hspace{5mm}
         \begin{minipage}{64mm}
        \begin{center}
       \unitlength=2mm
      \special{epsfile=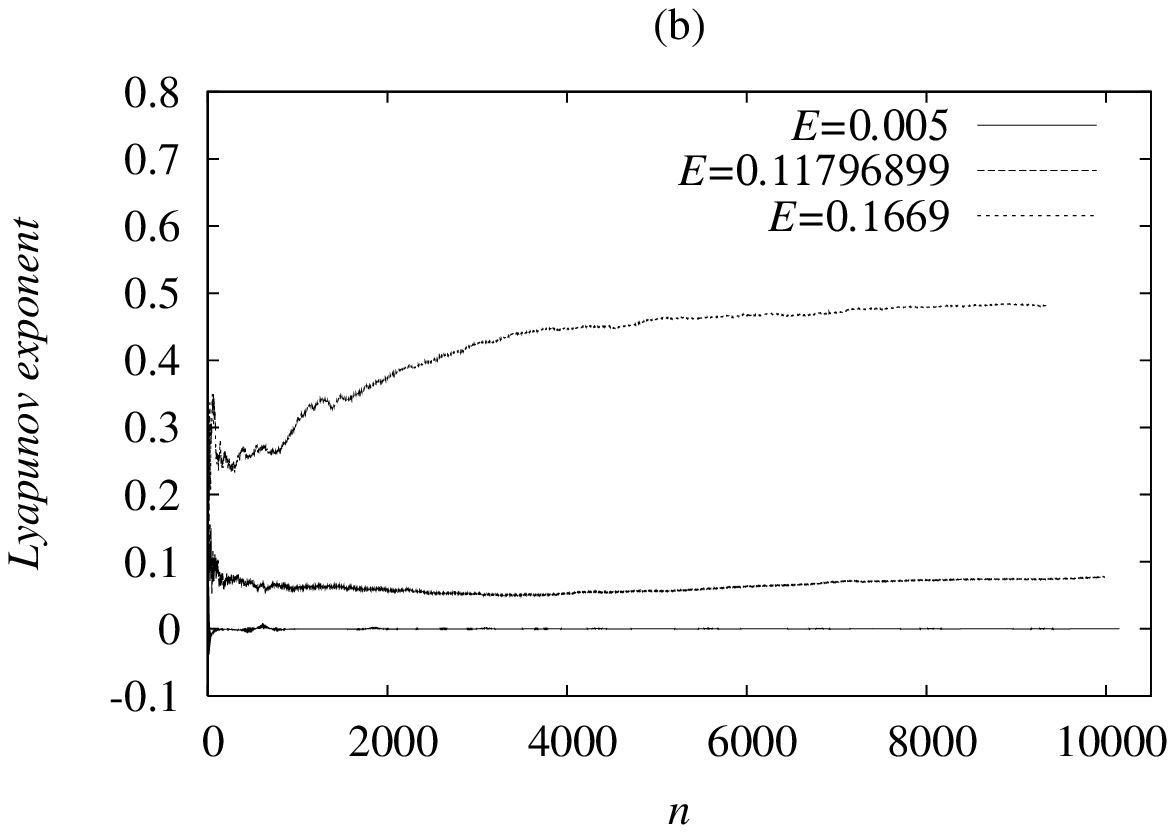 hscale=0.55 vscale=0.56}
     \vspace{53mm}
   \end{center}
  \end{minipage}
       \caption{(a) Time dependence of the absolute value of the correlation coefficient corresponding to Figs. 8(a)-(c). 
(b) Calculation of the Lyapunov exponent as a function of $n$.}
 \end{center}
\end{figure}

\begin{figure}[htb]
 \begin{center}
  \begin{minipage}{60mm}
   \begin{center}
    \unitlength=2mm
     \special{epsfile=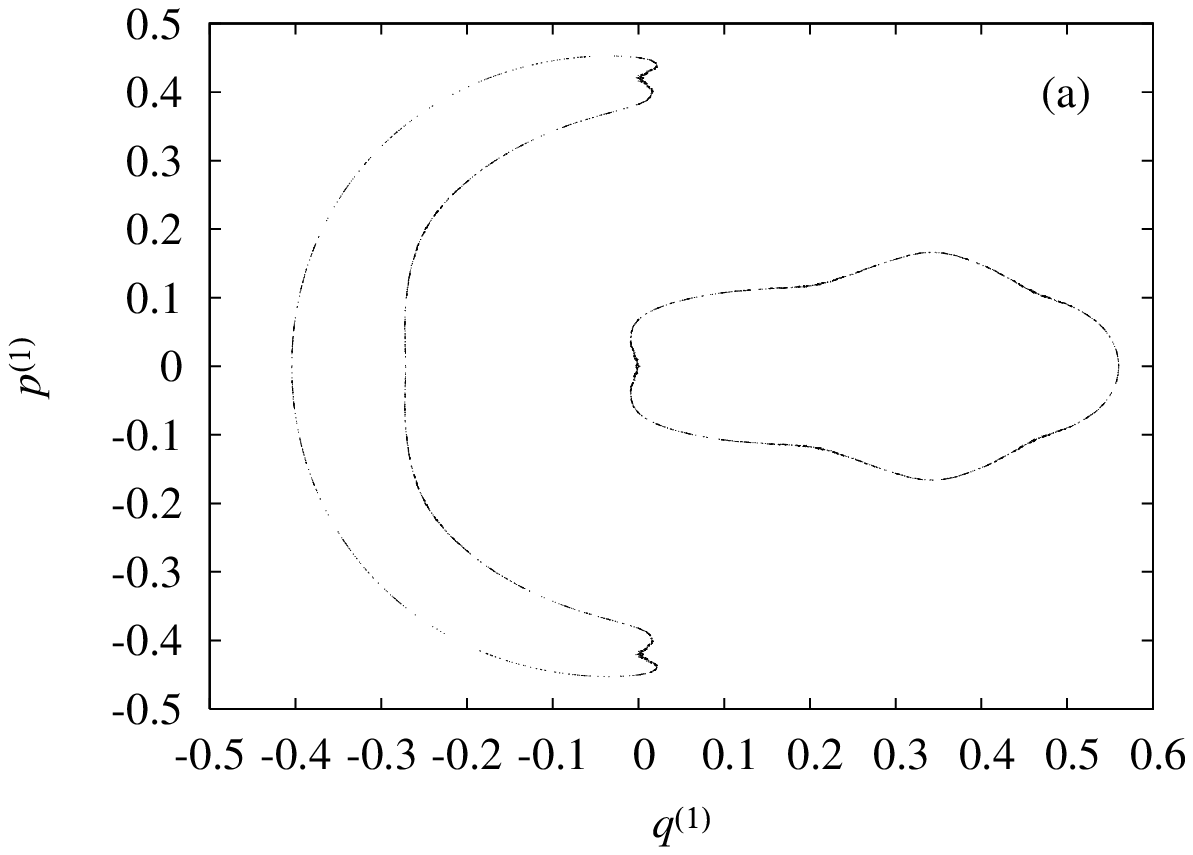 hscale=0.52 vscale=0.56}
      \vspace{53mm}
        \end{center}
         \end{minipage}
          \hspace{5mm}
         \begin{minipage}{60mm}
        \begin{center}
       \unitlength=2mm
      \special{epsfile=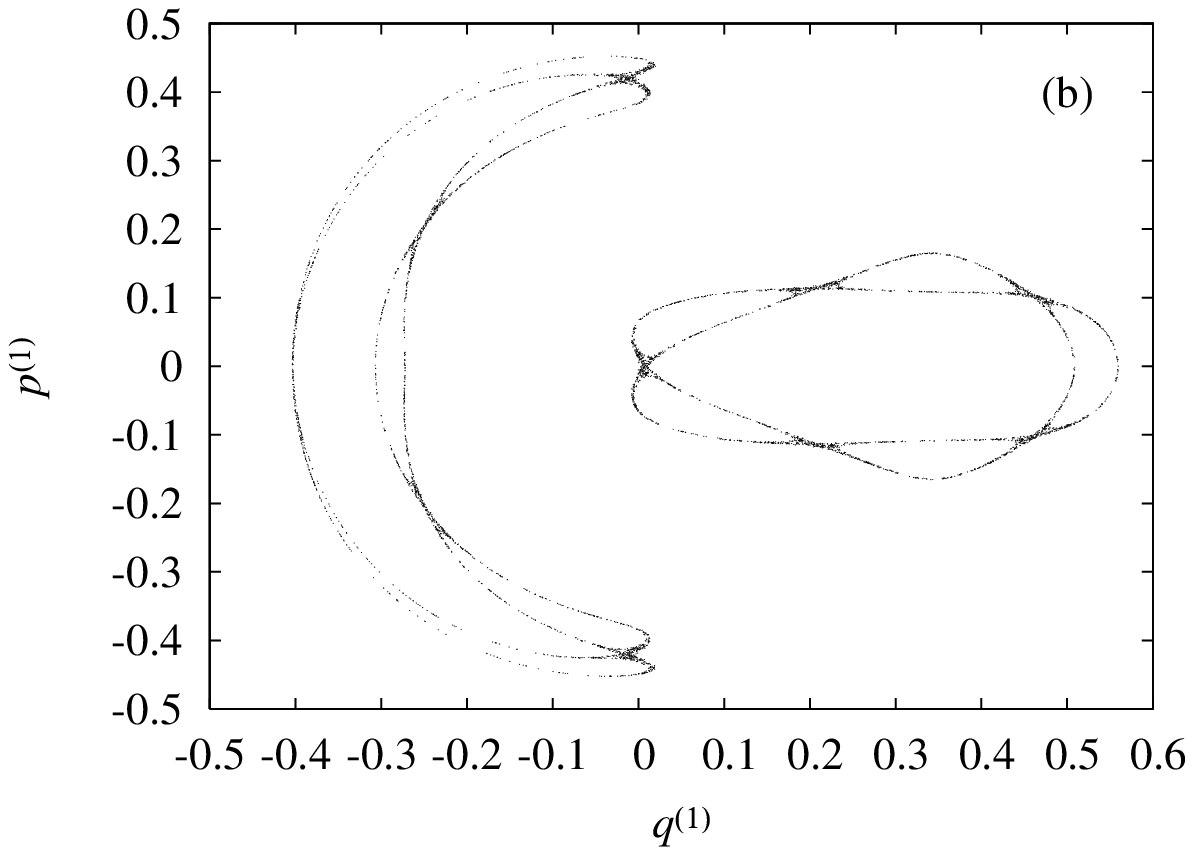 hscale=0.52 vscale=0.56}
     \vspace{53mm}
   \end{center}
  \end{minipage}
       \caption{The unstable orbit as shown in Fig. 7(b) is composed of orbit (a), which exists in the interval $n=0$ to $n=4900$, and orbit (b), which exists in the interval $n=4901$ to $n=9990$.}
 \end{center}
\end{figure}

Next, we describe an advantage that the correlation coefficient has when compared with the Lyapunov exponent. The unstable orbit of Fig. 7(b) is divided into two orbits, as depicted in Figs. 10(a)-(b). 
The absolute value of the correlation coefficient for the orbit in Fig. 7(b) is represented by the short dashed line corresponding to $E=0.11796899$ in Fig. 9(a). 
The value of $|C_{R_{m}r_{m}}|$ markedly varies in the region around $m\approx 2450$ as seen in Fig. 9(a). 
The corresponding values for the Lyapunov exponent are represented by the short dashed line in Fig. 9(b). As can be seen in those figures, there is a clear change in the absolute value of the correlation coefficient between the sections corresponding to the orbit of Fig. 10(a) and the orbit of Fig. 10(b).  
The Lyapunov exponent calculation shows the value does not vary markedly with increasing time, as illustrated in Fig. 9(b). 
In this case, the differing orbits of the two time sections shown in Figs. 10(a)-(b) cannot be distinguished clearly. 
This suggest a possible advantage of the correlation coefficient over the Lyapunov exponent. 
The numerical results discussed here show that it is possible to use $|C_{R_{m}r_{m}}|$ to sensitively detect even slight changes of orbit, something that the Lyapunov exponent is unable to do. 

\clearpage

\section{Symmetry violation and probability density distributions}

The main topic of this section is the determination of a relationship between a fluctuating property measuring symmetry violation and nonintegrability. 
To begin with, we define a new variable $O_{m}=R_{m}\times r_{m}$. 
When the perturbation interaction is sufficiently small, as in Fig. 4(a), this quantity is stationary with a value of ${O_{m}}\approx 1.0$ as shown in Fig. 11(a). 
This corresponds to the case where there is symmetry between $R_{m}$ and $r_{m}$ and $|C_{R_{m}r_{m}}|\approx 1.0$. 
This suggests the following equation for the time evolution $m$ when the system is completely integrable.
\begin{eqnarray}
\frac{dO_{m}}{dm}&=&0.
\end{eqnarray}

\begin{figure}[htb]
 \begin{center}
  \begin{minipage}{62mm}
   \begin{center}
    \unitlength=2mm
     \special{epsfile=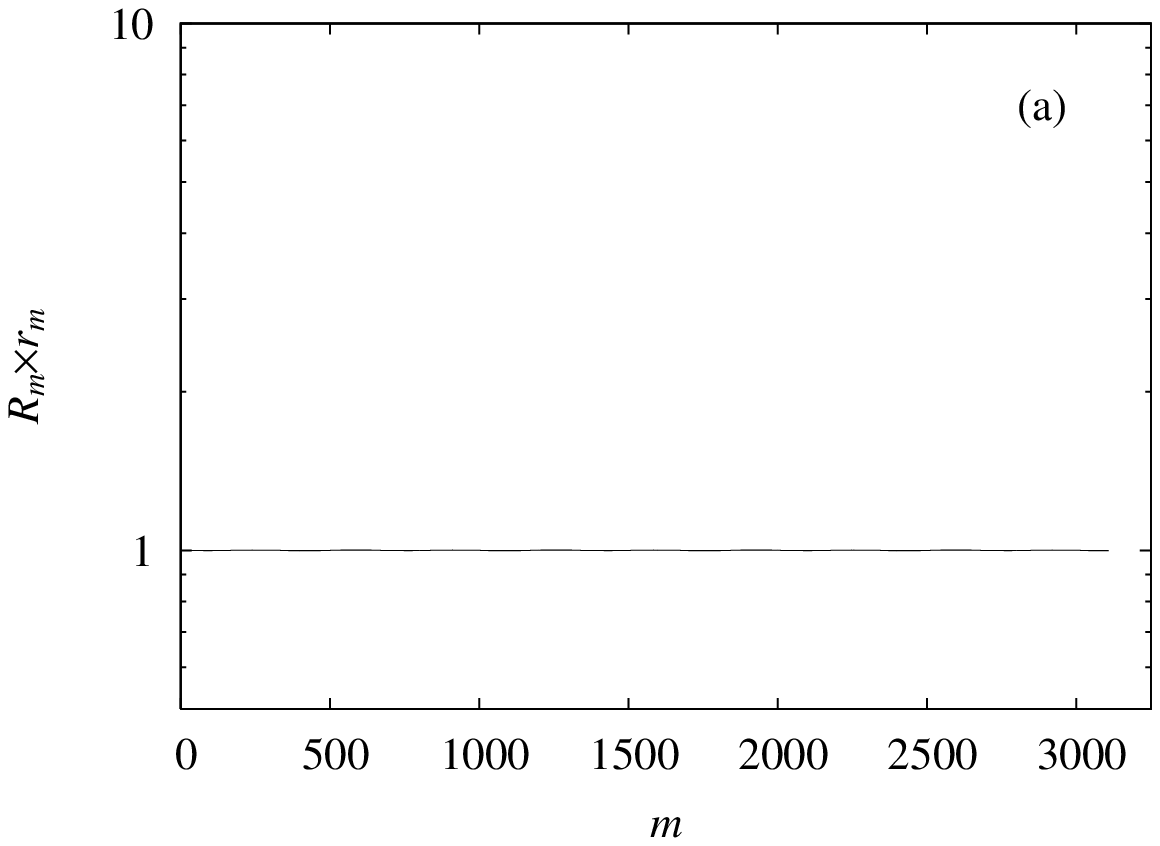 hscale=0.55 vscale=0.55}
      \vspace{51mm}
        \end{center}
         \end{minipage}
          \hspace{8mm}
         \begin{minipage}{62mm}
        \begin{center}
       \unitlength=2mm
      \special{epsfile=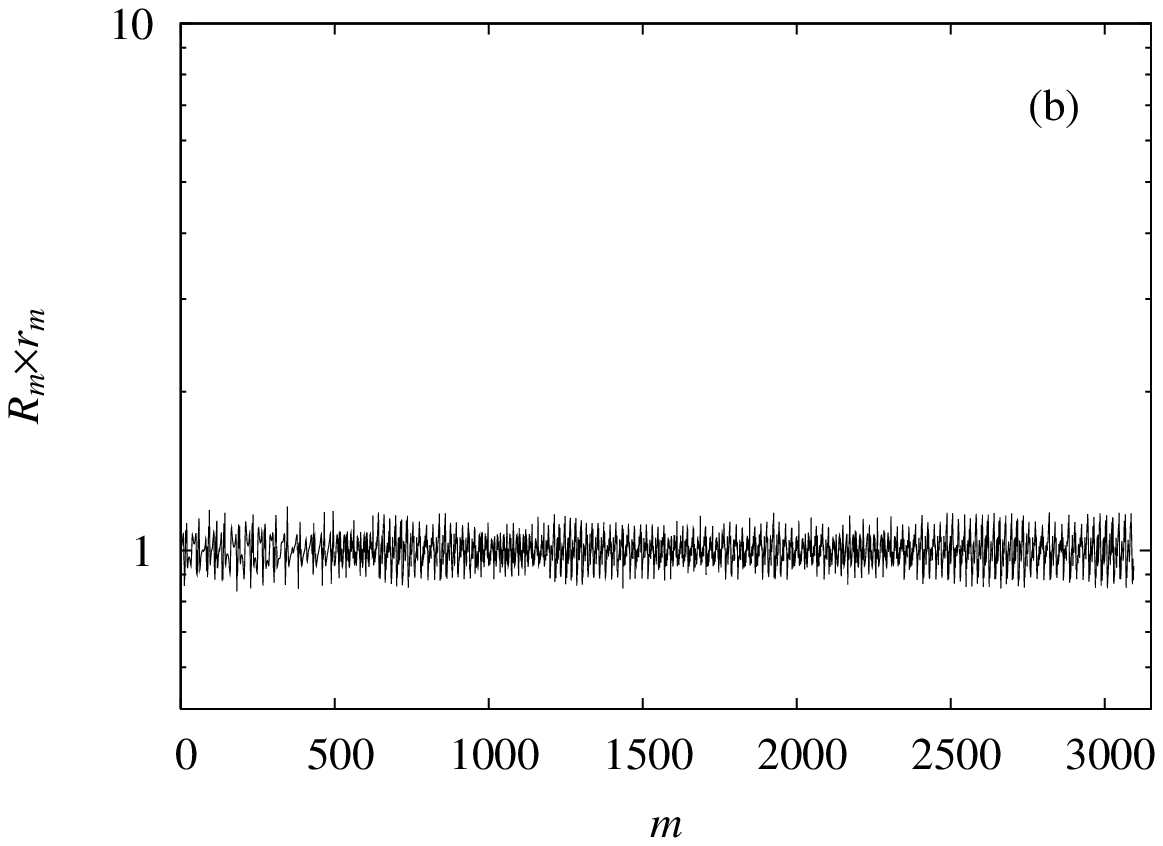 hscale=0.55 vscale=0.55}
     \vspace{51mm}
   \end{center}
  \end{minipage}
          \hspace{8mm}
         \begin{minipage}{62mm}
        \begin{center}
       \unitlength=2mm
      \special{epsfile=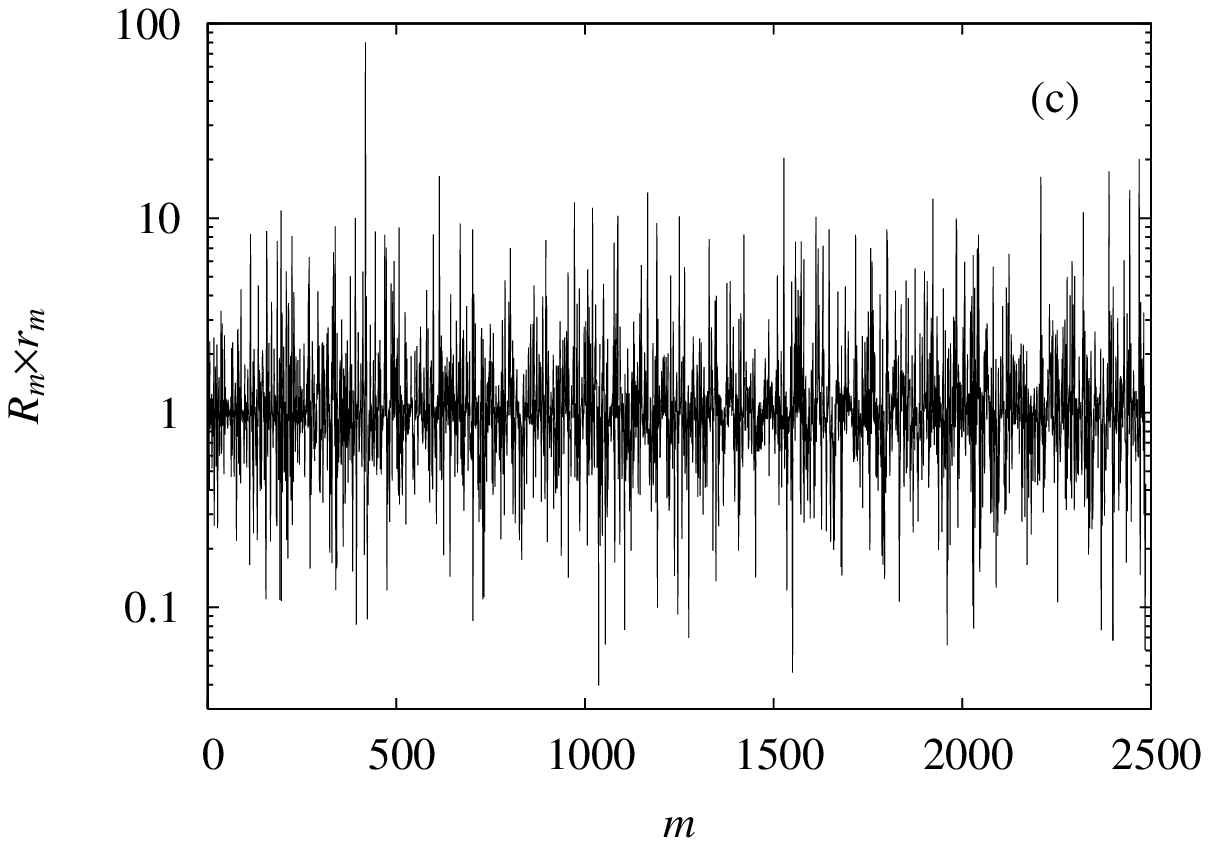 hscale=0.55 vscale=0.55}
     \vspace{51mm}
   \end{center}
  \end{minipage}
 \end{center}
       \caption{Computation of the fluctuation property $O_{m}=R_{m}\times r_{m}$ for perturbation interaction values (a) $V=-0.009$, (b) $V=-0.045$ and (c) $V=-0.110$.}
         \end{figure}

\begin{figure}[htb]
 \begin{center}
  \begin{minipage}{62mm}
   \begin{center}
    \unitlength=2mm
     \special{epsfile=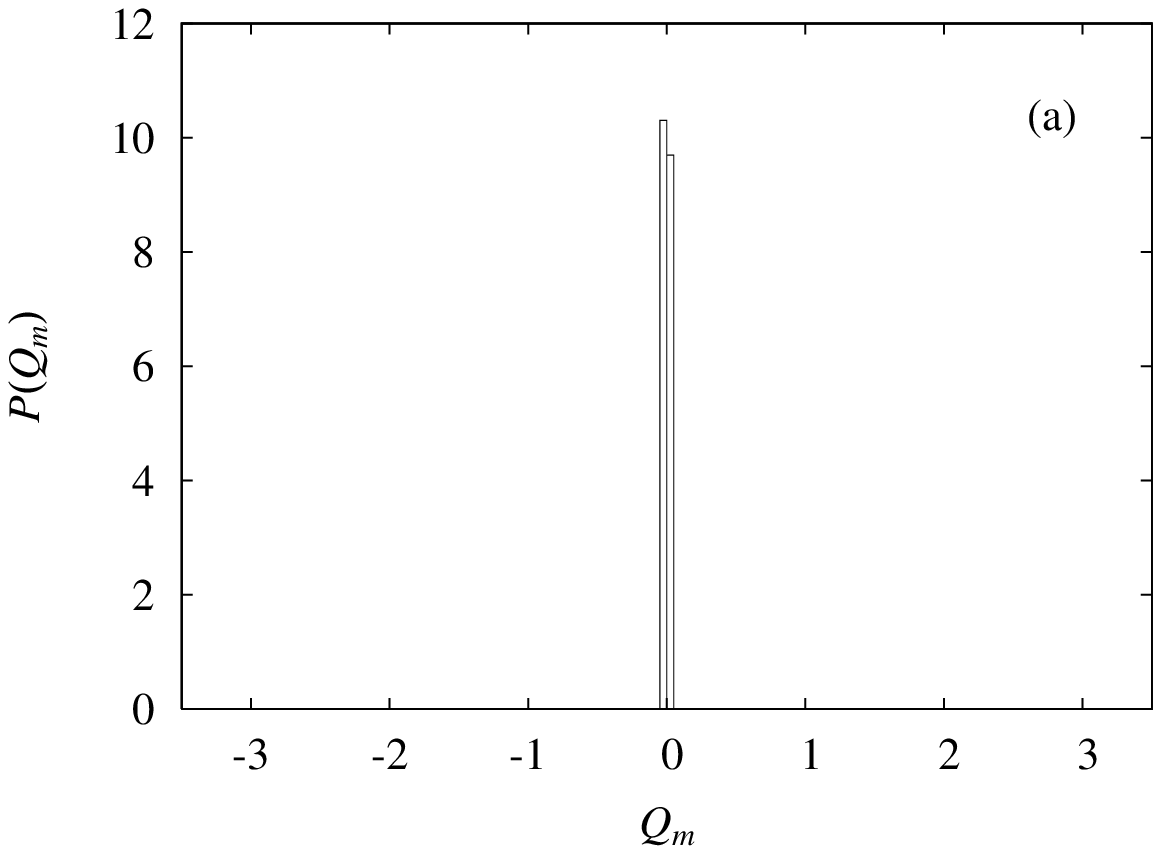 hscale=0.55 vscale=0.55}
      \vspace{51mm}
        \end{center}
         \end{minipage}
          \hspace{8mm}
         \begin{minipage}{62mm}
        \begin{center}
       \unitlength=2mm
      \special{epsfile=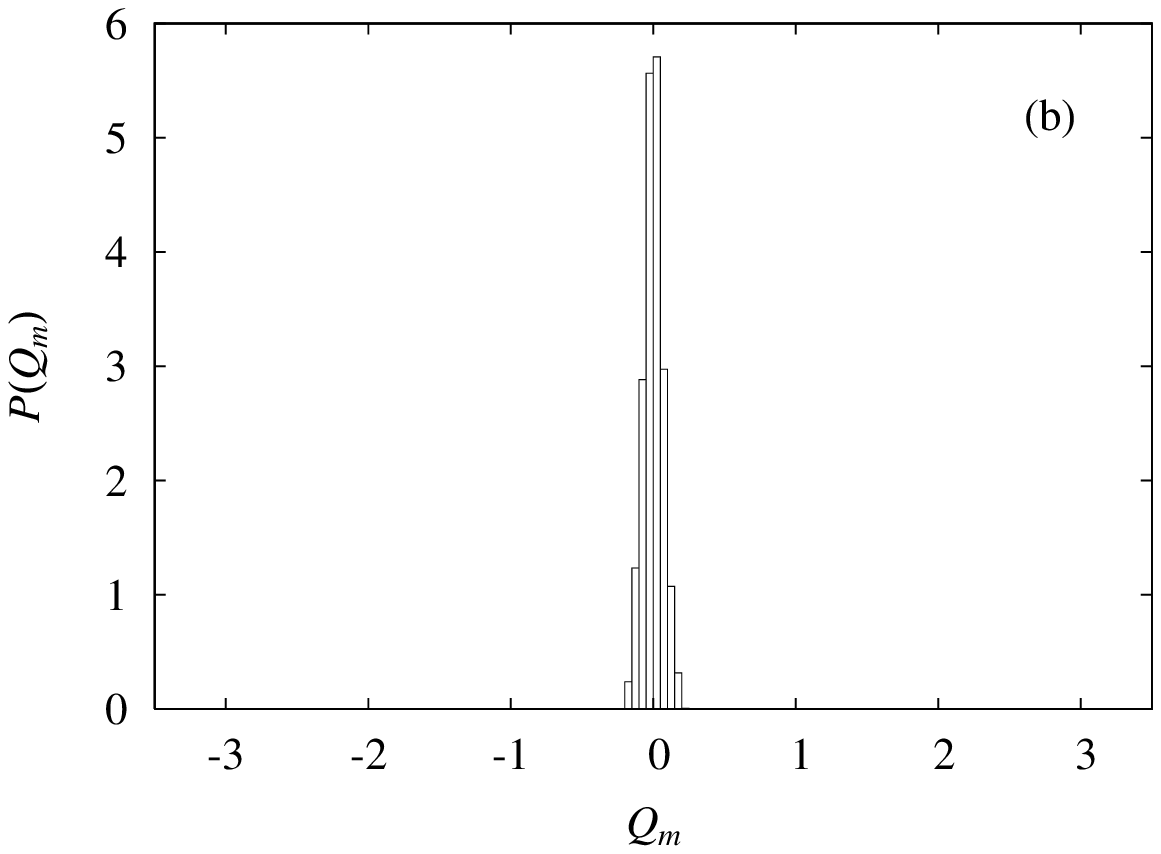 hscale=0.52 vscale=0.56}
     \vspace{51mm}
   \end{center}
  \end{minipage}
          \hspace{8mm}
         \begin{minipage}{62mm}
        \begin{center}
       \unitlength=2mm
      \special{epsfile=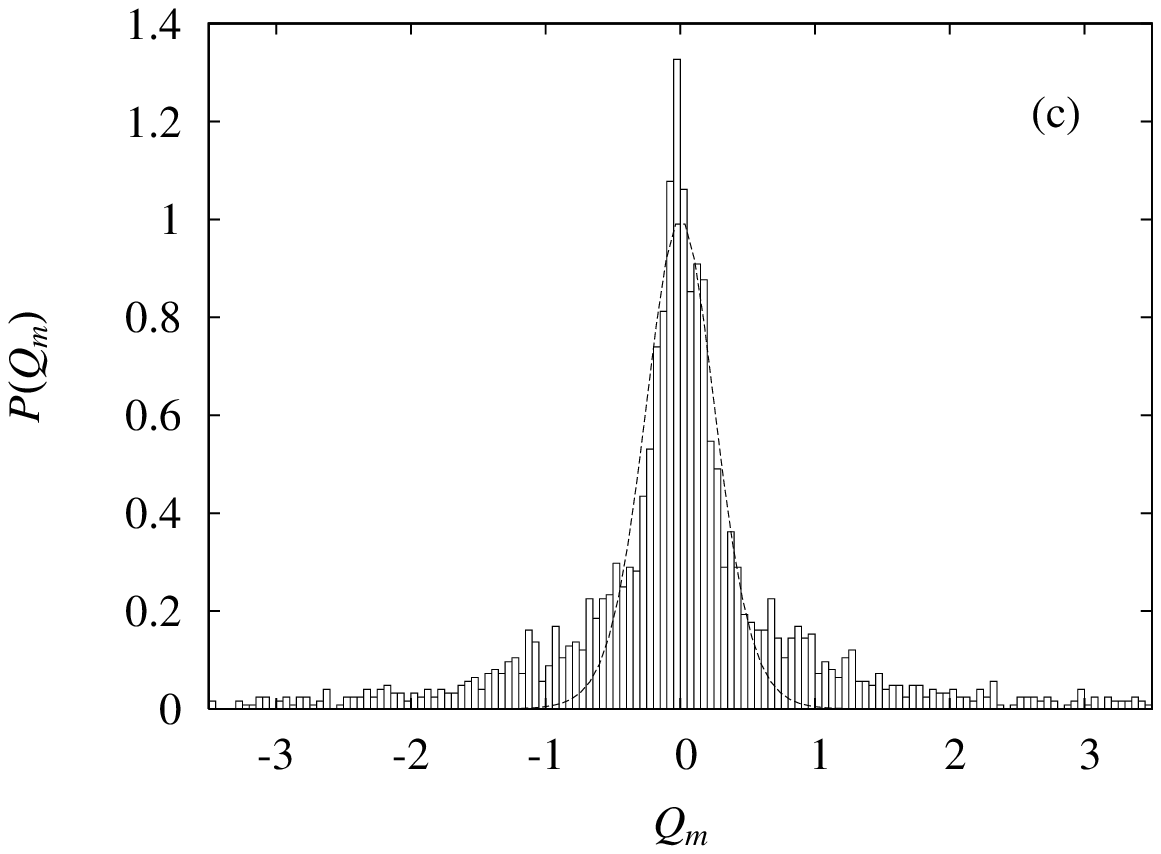 hscale=0.55 vscale=0.55}
     \vspace{51mm}
   \end{center}
  \end{minipage}
 \end{center}
       \caption{Probability density distribution of $Q_{m}$. The width of the histogram $dQ_{m}$ is $0.05$. (a) $V=-0.009$, (b) $V=-0.045$, and (c) $V=-0.110$. The dotted line is the function with the paremeter $\alpha =14.5$ in Eq. (6.3).}
         \end{figure}
In the case of Figs. 4(b)-(c), the balance between the elongation rate and contraction rate of the phase-space point spacings is broken, 
and this is confirmed in Figs. 11(b)-(c) as a fluctuation where $\frac{dO_{m}}{dm}\neq 0$. 
The statistical properties of this fluctuation are now discussed. 
To handle the logarithmic behavior between the elongation rate and contraction rates, and hence of $O_{m}$, let us introduce the variable $Q_{m}$, where
\begin{eqnarray}
Q_{m} = \left\{
\begin{array}{ll}
O_{m}-1 &, \quad O_{m} \geq 1\\
1-\frac{1}{O_{m}} &, \quad O_{m} < 1\hspace{1mm}.\\
\end{array}
\right.
\end{eqnarray}
For example, the values $O_{1}=2.0$, $O_{2}=5.0$, $O_{3}=10.0$, $O_{4}=0.5$, $O_{5}=0.2$, and $O_{6}=0.1$ correspond to $Q_{1}=1.0$, $Q_{2}=4.0$, $Q_{3}=9.0$, $Q_{4}=-1.0$, $Q_{5}=-4.0$, and $Q_{6}=-9.0$.
This variable behaves in a manner similar to $O_{m}$ for symmetry violations. 

We now investigate the distribution of $Q_{m}$ with increasing nonintegrability. 
Figures. 12(a)-(c) show the probability density $P(Q_{m})$ of finding for the components a value between $Q_{m}$ and $Q_{m}+dQ_{m}$ for each interaction value. 
The probability density of $Q_{m}$ behaves as a delta function when the system is integrable, as shown in Fig. 12(a). 
On the other hand, it becomes a non-Gaussian function as follows, such as that shown in Fig. 12(c) when the nonintegrability is intensified. 
\begin{eqnarray}
P(Q_{m}) &=& \frac{1}{({Q_{m}}^{2}+1)^{\frac{1+\alpha}{2}}}
\end{eqnarray}
Where the parameter $\alpha$ is fitted by the general least squares procedure\cite{rf:6}. 

In the case of the H\'{e}non-Heiles system at an energy of 0.1669, $P(Q_{m})$ also shows a function in Eq. (6.3), as can be seen in Fig. 14(c). 
This corresponds to the fluctuation property $O_{m}$ in Fig. 13(c).

Note that the statistical property of $O_{m}$ changes from a delta function to a function in Eq. (6.3) with increasing perturbation strength. 
This highlights the importance of $O_{m}$ as a fluctuating measure of non-integrability in the perturbed harmonic oscillator systems.

\begin{figure}[htb]
 \begin{center}
  \begin{minipage}{62mm}
   \begin{center}
    \unitlength=2mm
     \special{epsfile=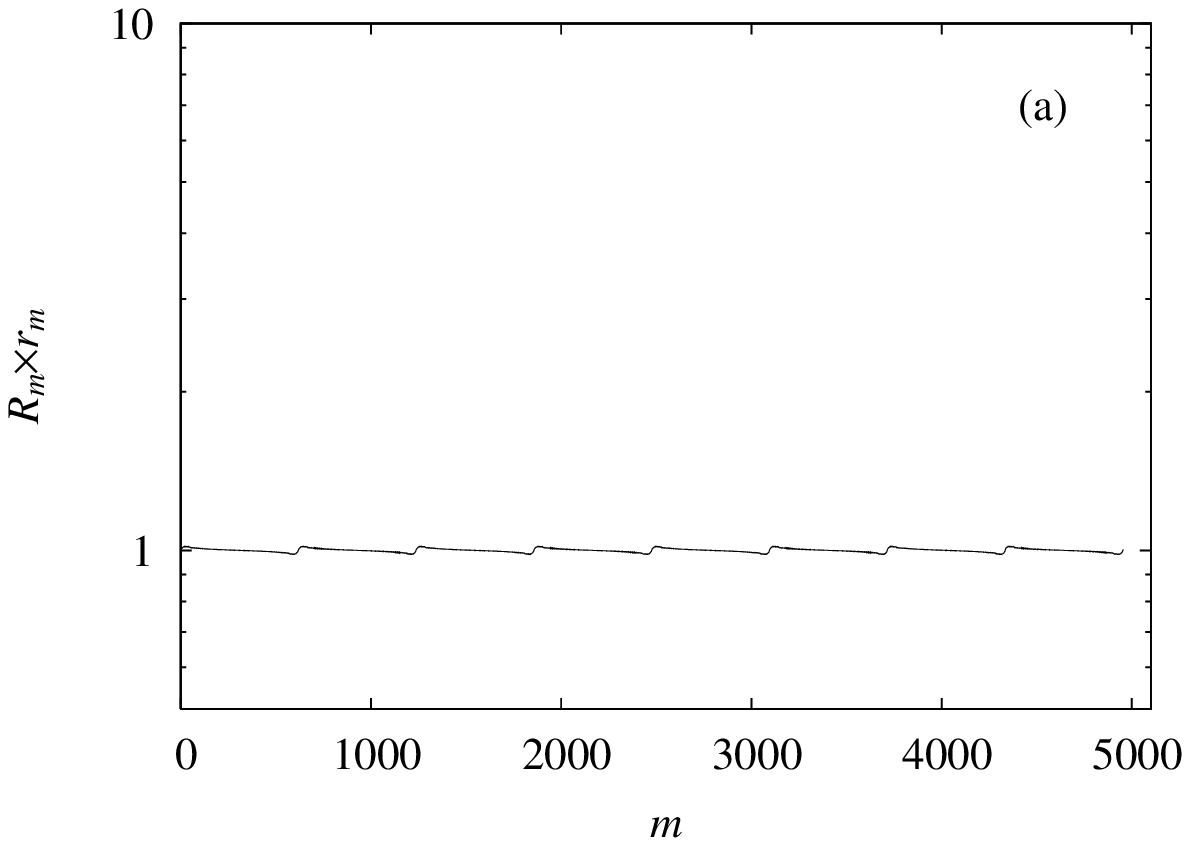 hscale=0.55 vscale=0.55}
      \vspace{51mm}
        \end{center}
         \end{minipage}
          \hspace{8mm}
         \begin{minipage}{62mm}
        \begin{center}
       \unitlength=2mm
      \special{epsfile=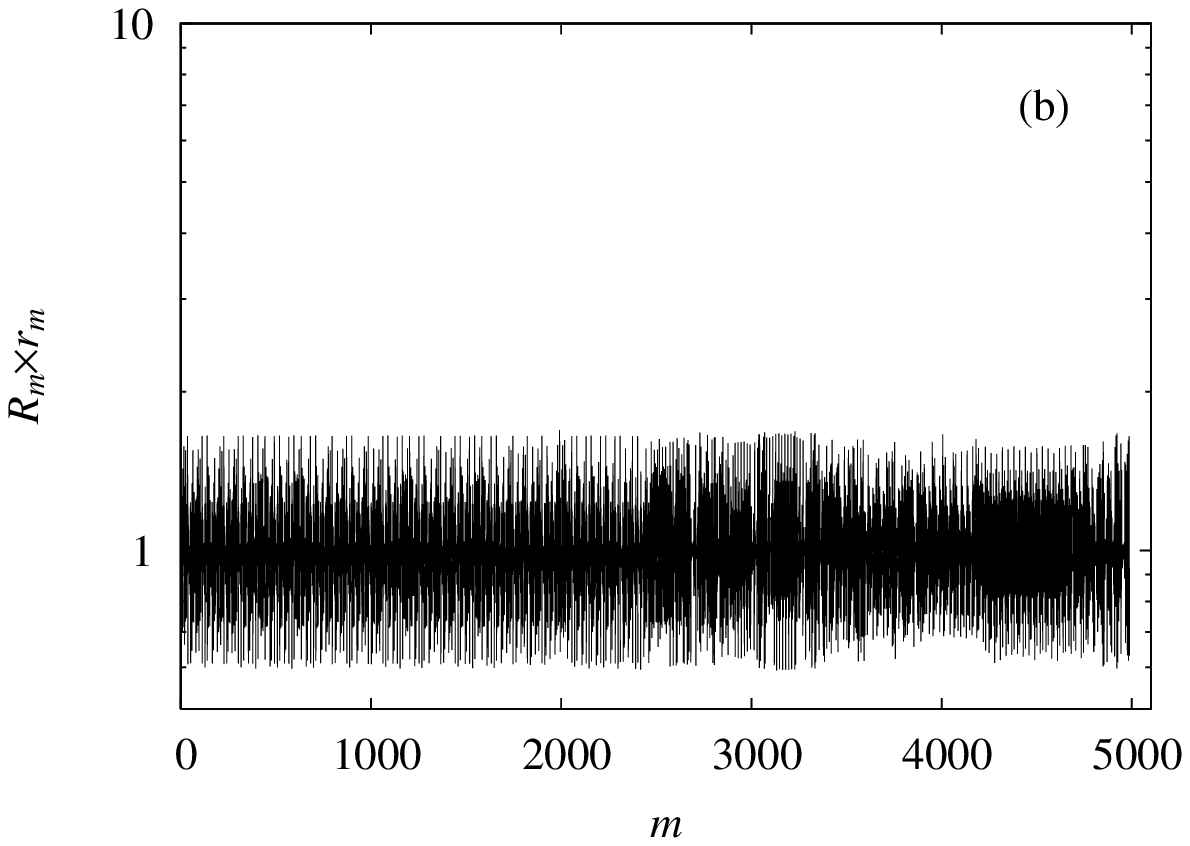 hscale=0.55 vscale=0.55}
     \vspace{51mm}
   \end{center}
  \end{minipage}
          \hspace{8mm}
         \begin{minipage}{62mm}
        \begin{center}
       \unitlength=2mm
      \special{epsfile=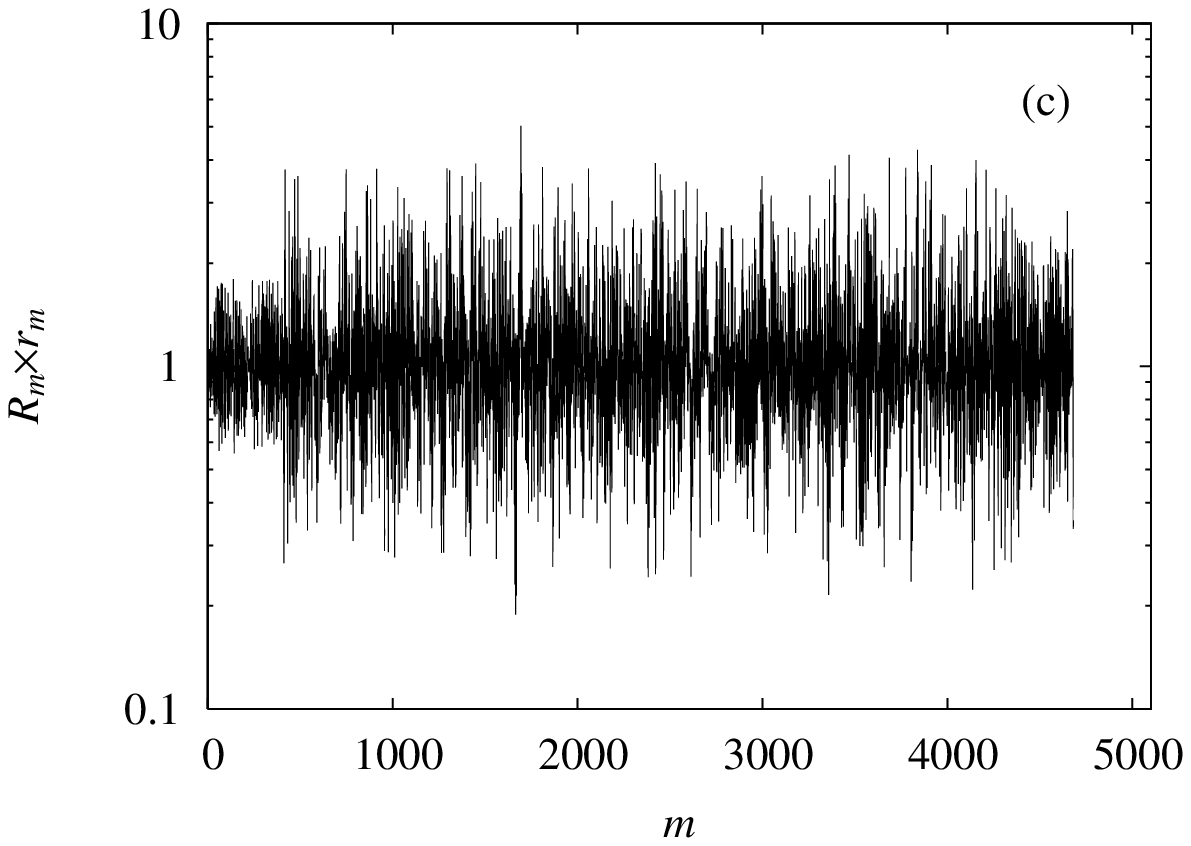 hscale=0.55 vscale=0.55}
     \vspace{51mm}
   \end{center}
  \end{minipage}
 \end{center}
       \caption{Computation of the fluctuation property $O_{m}=R_{m}\times r_{m}$ for the H\'{e}non-Heiles Hamiltonian with (a) $E=0.005$, (b) $E=0.11796899$, and (c) $E=0.1669$.}
         \end{figure}

\begin{figure}[htb]
 \begin{center}
  \begin{minipage}{62mm}
   \begin{center}
    \unitlength=2mm
     \special{epsfile=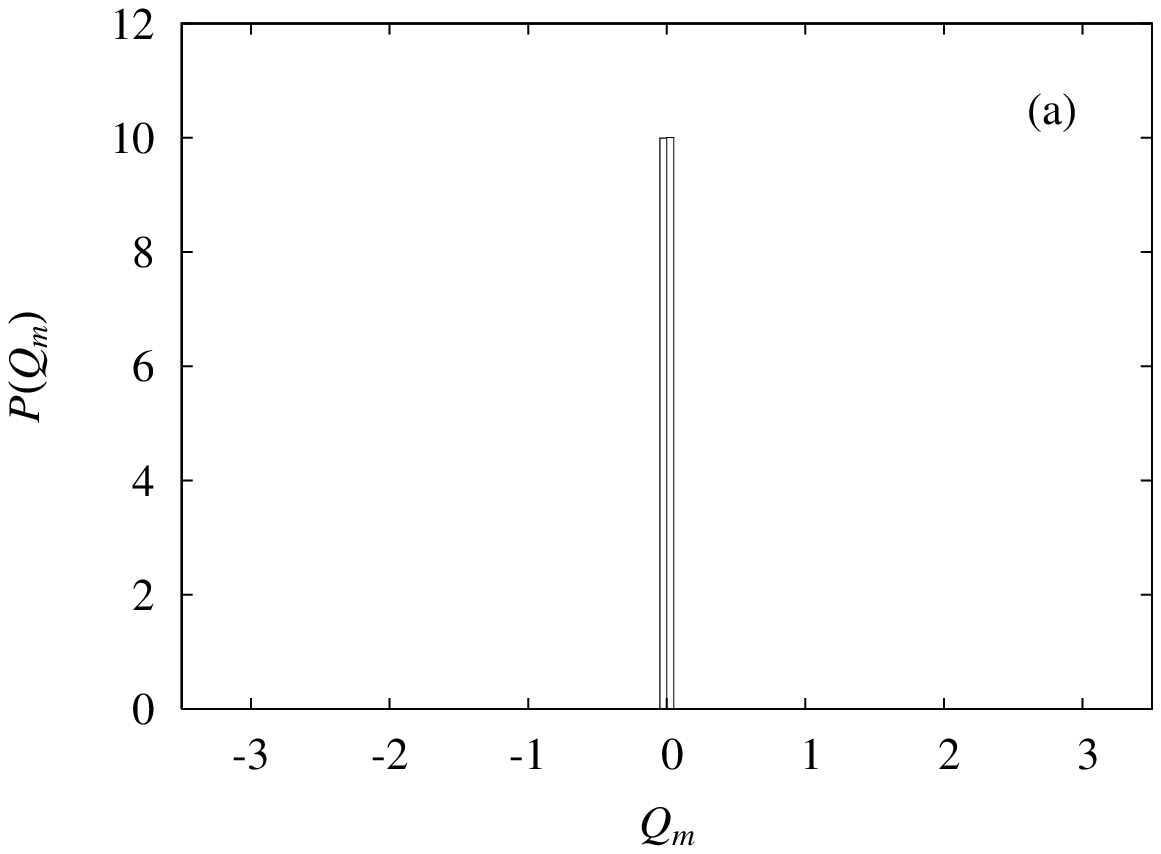 hscale=0.55 vscale=0.55}
      \vspace{51mm}
        \end{center}
         \end{minipage}
          \hspace{8mm}
         \begin{minipage}{62mm}
        \begin{center}
       \unitlength=2mm
      \special{epsfile=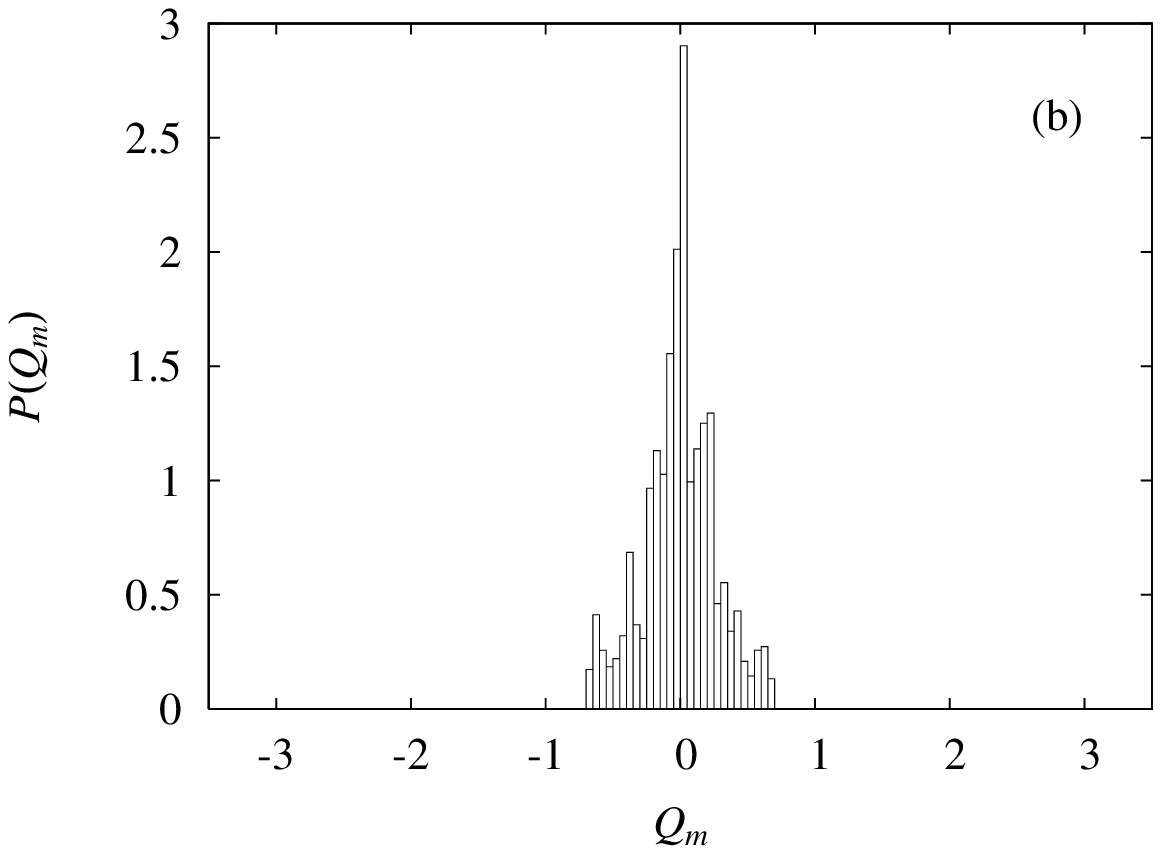 hscale=0.55 vscale=0.55}
     \vspace{51mm}
   \end{center}
  \end{minipage}
          \hspace{8mm}
         \begin{minipage}{62mm}
        \begin{center}
       \unitlength=2mm
      \special{epsfile=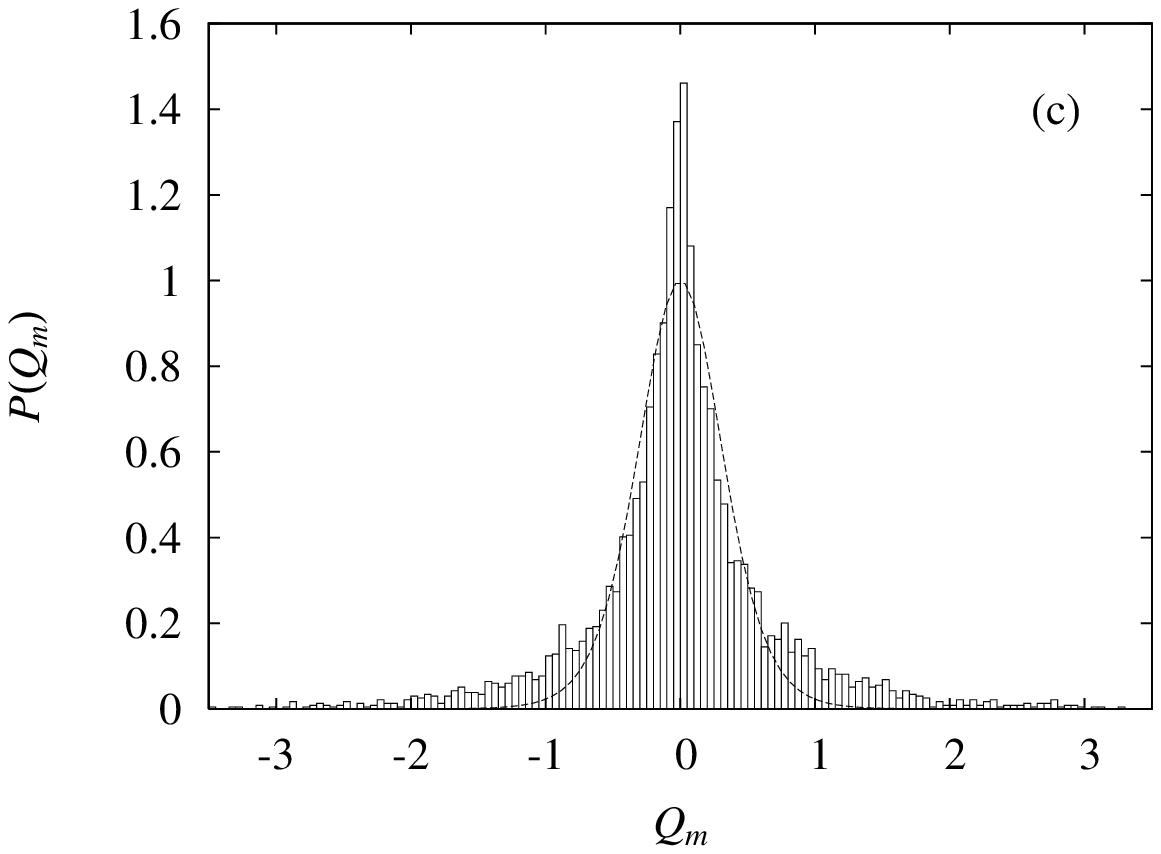 hscale=0.55 vscale=0.55}
     \vspace{51mm}
   \end{center}
  \end{minipage}
 \end{center}
       \caption{Probability density distribution of $Q_{m}$. The width of the histogram $dQ_{m}$ is $0.05$. (a) $E=-0.005$, (b) $E=0.11796899$, and (c) $E=0.1669$. The dotted line is the function with the paremeter $\alpha =9.9$ in Eq. (6.3).}
         \end{figure}

\section{Conclusions}

We conclude that this analysis provides us with a new method of detecting nonintegrabilities based on the degree of symmetry violation between $R_{m}$ and $r_{m}$.
It was demonstrated that slight changes in the properties of unstable orbits, which cannot be distinguished using the Lyapunov exponent, could be detected when using the absolute value of the correlation coefficient.

We also verified numerically that the breaking of symmetry caused the probability density for $Q_{m}$ to become the function in Eq. (6.3). 
It is hoped that a distribution function that describes the symmetry violation may be found in the future. 
Further research on the statistical principle that determines the probability distribution function for generating symmetry violation is needed; thus, statistical mechanics which can describe chaotic motion that breaks the symmetry based on this principle should be constructed.

\section*{Acknowledgements}
I would like to express my heartfelt thanks to Dr. Fumihiko SAKATA for stimulating discussions.

%\appendix
%\section{First Appendix} %Empty argument \section{} yields `Appendix'. 
%
%\section{Second Appendix}


\begin{thebibliography}{99}
\bibliography{sym_levy_E2}% Produces the bibliography via BibTeX.

\bibitem{rf:1} A. J. Lichtenberg and M. A. Lieberman,
{\it Regular and Stochastic Motion}, Applied Mathematical Sciences, {\bf 38}
(Springer, Berlin, 1983).
\bibitem{rf:2} E. Ott, {\it Chaos in Dynamical Systems}, (Cambridge Univ, 1993)
\bibitem{rf:3} S. Y. Li, A. Klein, and R. M. Dreizler,
J. Math. Phys. {\bf 11}, (1970), 975.
\bibitem{rf:4} F. Sakata, T. Marumori, Y. Hashimoto, H. Tsukuma, Y. Yamamoto, J. Terasaki,
Y. Iwasawa, and H. Itabashi,  {\it Springer Proceedings in Physics}, {\bf 58}, (Springer-Verlag, Berlin, Heidelberg, 1991), 187.
\bibitem{rf:5} Rubin. H. Landau and M. J. Paez,
 {\it Computational Physics, Problem Solving With Computers} (Wiley-Interscience, NY, 1997). 
\bibitem{rf:6} W. H. Press, S. A. Teukolsky, W. T. Vetterling, and B. P. Flannery,
{\it Numerical Recipes in C}, 2nd ed. (Cambridge, 1992).
\end{thebibliography}
\end{document}